\documentclass[aps,pre,showpacs,superscriptaddress]{revtex4-1}
\usepackage[english]{babel}
\usepackage{graphicx}
\usepackage[caption=false]{subfig}

\usepackage[utf8]{inputenc}
\usepackage{amssymb}
\usepackage{amsmath}
\usepackage{siunitx}
\usepackage{hyperref}
\hypersetup{colorlinks=true}
\newcommand{\nm}[1]{\SI{#1}{\nano\meter}}

\newcommand{\Lc}{L_c}								
\newcommand{\cs}{c_s}
\newcommand{\En}{\mathcal{E}}						        
\newcommand{\Free}{\mathcal{F}}						
\DeclareMathOperator{\Tw}{T\!w} 								
\DeclareMathOperator{\Wr}{W\!r} 								
\DeclareMathOperator{\Lk}{Lk} 								
\newcommand{\pb}{P_b}								
\newcommand{\pc}{P_c}								
\newcommand{\pcef}{P_c^{\textrm{ren}}}					
\DeclareMathOperator{\kt}{k_BT}								
\newcommand{\Lp}{L_p}									
\DeclareMathOperator{\lamp}{W_p}
\DeclareMathOperator{\lamm}{W_m}

\DeclareMathOperator{\tw}{t\!w} 								
\newcommand{\wri}{\omega} 							
\DeclareMathOperator{\lk}{lk} 								
\newcommand{\neff}{q_{\textrm{eff}}} 					
\newcommand{\free}{\mathfrak{f}}						
\newcommand{\lp}{l_p}
\DeclareMathOperator{\Wrl}{W\!r_{\textrm{loop}}} 				
\DeclareMathOperator{\Wrll}{W\!r_{\textrm{loop}}^1}				
\newcommand{\Enl}{\mathcal{E}_{\textrm{loop}}}			
\newcommand{\Ll}{L_{\textrm{loop}}}					

\DeclareMathOperator{\wrl}{\omega_{\textrm{loop}}}				

\DeclareMathOperator{\rt}{\rho_{\textrm{tail}}} 					
\DeclareMathOperator{\rp}{\rho_{\textrm{pl}}} 					
\DeclareMathOperator{\tws}{t\!w_{\textrm{str}}}					
\DeclareMathOperator{\lks}{lk_{\textrm{str}}}							
\newcommand{\ftwt}{\mathfrak{f}_{\textrm{tw}}^{\textrm{t}}}		
\newcommand{\ftwp}{\mathfrak{f}_{\textrm{tw}}^{\textrm{str}}}	
\newcommand{\ftw}{\mathfrak{f}_{\textrm{tw}}}					
\newcommand{\eb}{\mathfrak{e}_{\textrm{bend}}}				
\newcommand{\fb}{\mathfrak{f}_{\textrm{bend}}}					
\newcommand{\epl}{\mathfrak{e}_{\textrm{eplect}}}
\newcommand{\fpl}{\mathfrak{f}_{\textrm{plect}}}
\newcommand{\gpl}{g_{\textrm{plect}}}
\newcommand{\eel}{\mathfrak{e}_{\textrm{el}}^0}
\newcommand{\fel}{\mathfrak{f}_{\textrm{el}}}
\newcommand{\fstr}{\mathfrak{f}_{\textrm{strand}}}

\DeclareMathOperator{\sech}{sech}
\newcommand{\ftail}{\mathfrak{f}_{\textrm{tail}}}
\newcommand{\gtail}{g_{\textrm{tail}}}
\newcommand{\dg}{\Delta g}

\newcommand{\wdt}{\omega^{\textrm{th}}_{\textrm{tail}}}
\DeclareSIUnit\Molar{M}
\newcommand{\vev}[1]{ \left\langle #1 \right\rangle}
\newcommand{\de}{\textrm{d}}
\renewcommand{\vec}[1]{\mathbf{#1}} %
\begin{document}
\title{Multi-plectoneme phase of double-stranded DNA under torsion}
\author{Marc \surname{Emanuel}}
\affiliation{Instituut Lorentz voor de theoretische %
natuurkunde, Universiteit Leiden,%
P.O. Box 9506, NL-2300 RA Leiden, The Netherlands}
\affiliation{Institute of Complex Systems II, Forschungszentrum J\"ulich, J\"ulich 52425, Germany,}
\affiliation{Delft University of Technology, Department of Bionanoscience, Kavli Institute of nanoscience, Lorentzweg 1, 2628CJ Delft, The Netherlands,}
\author{Giovanni \surname{Lanzani}}
\author{Helmut \surname{Schiessel}}
\affiliation{Instituut Lorentz voor de theoretische %
natuurkunde, Universiteit Leiden,%
P.O. Box 9506, NL-2300 RA Leiden, The Netherlands}
\pacs{64.70.km,87.10.Ca,87.15.ad}
\begin{abstract}
  We use the worm-like chain model to study
  supercoiling of DNA under tension and torque. The model reproduces
  experimental data for a broad range of forces, salt concentrations and
  contour lengths.
  We find a plane of first order phase transitions ending in a smeared out line of critical points, the multi-plectoneme phase, which is characterized by a fast twist mediated diffusion of plectonemes and a torque that rises after plectoneme formation with increasing linking number. The discovery of this new phase at the same time resolves the discrepancies between existing models and experiment.
\end{abstract}
\maketitle
\section{\label{sec:Introduction}Introduction}

\begin{figure}[htbf] \centering
  \includegraphics[width=\linewidth]{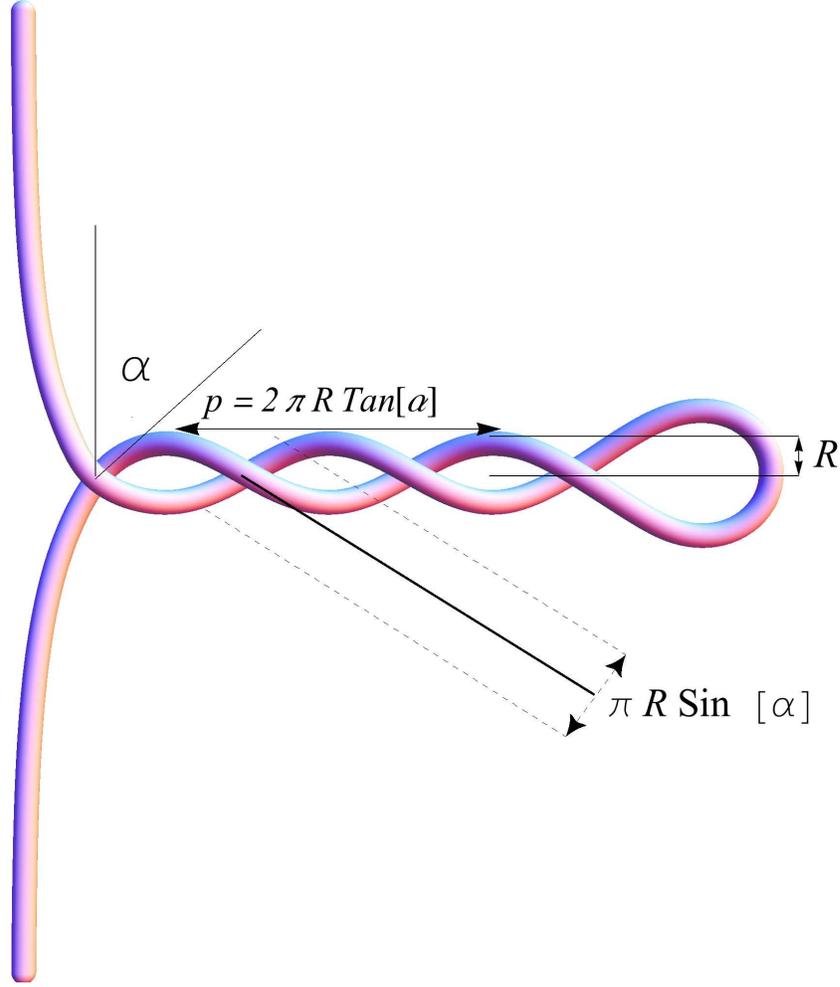}
  \caption{A plectoneme with plectoneme radius $R$, plectoneme angle $\alpha$ and pitch $p$. The standard deviation of the fluctuation channel in the ``pitch-direction'' is $\pi R \sin(\alpha)$}
  \label{fig:plectoneme}
\end{figure}

The behavior of double stranded DNA (dsDNA) under tension and torsion plays an important role in the
transcription and replication of our genetic code. The DNA present in
a single human cell is long enough to outdo most of us in
height: yet it is confined in nuclei with diameters in the micron range, orders of magnitude smaller than the chain would have in a theta solvent.  One
of the ingredients in the compactification of DNA in bacteria is supercoiling, where
torsional stress results in the formation of plectonemes: loops in the molecule with the two halves of the molecule coiled around each other, like an old-fashioned telephone cord (Fig.~\ref{fig:plectoneme}). Since dsDNA forms in its relaxed state a right handed double helix it is
chiral, and a combination of torsion and tension comes automatically into play during
transcription and replication. Single molecule experiments\cite{Smith:1992} have been
instrumental in investigating the elastic properties of dsDNA. The force extension behavior of a freely rotating chain
can be described by modeling the molecule as an elastic rod in a thermal environment\cite{Marko:1995a,Odijk:1995}, with
all elastic strains are described by just two elastic moduli: the bending modulus, and a stretch modulus $S$ which due to its large value of $\sim 700-1300$ pN can safely be
omitted for tensions in the pN range. This worm like chain (WLC) model was shown\cite{Marko:1995a,Odijk:1995} to be a good
description over a large range of contour-lengths and tensions.

When experimental techniques made it possible to put at the same time a torque on the molecule\cite{Strick:1996}, adding the torsional degree of freedom with its linear elastic modulus to the WLC model results in a good description of the experimental data\cite{Moroz:1998} for torques somewhat lower than the classical buckling transition. After this buckling transition a growing plectoneme (Fig.~\ref{fig:plectoneme}) is thought to set the slope of the force extension curve. Many models have been constructed to predict some of the measurements but, as we will argue in this paper, are incomplete in their description of the thermal fluctuations. This led to some remarkable disagreement with experimental data. Furthermore an important feature of the phase diagram of torsionally stressed dsDNA remained uncovered.
The experimental setup that we compare the model with consist of a dsDNA molecule that has one end attached to a substrate, the other end attached to a bead. This bead is either super-paramagnetic (small ferromagnetic domains randomly oriented in a polystyrene sphere) or glass. In the first case the position is controlled by the gradient of a magnetic field hence the name magnetic tweezer, the second by laser-beams, the optical tweezer.  Making use of a net magnetic moment (or a specially crafted bead in the optical case) it is possible to control the rotation of the bead. The torque is not fixed in these measurements, but the average torque is extracted either from the rotation extension curves, or from the trap stiffness using specially crafted beads.

In this paper we will include a consistent description of thermal fluctuations, on the way adding some details to the common models, that remained somewhat hidden in the usual treatments. For notational convenience we scale all energies by $\kt$ unless explicitly mentioned. Forces will have the dimension of an inverse length and the two moduli introduced that of a length.

The setup of the paper is the following: in section~\ref{sec:1} we define the Hamiltonian using the contributions that are common to most models, on the way putting in some details that are not always appreciated. In section~\ref{sec:thermal} we will analyze in detail the influence of thermal fluctuations on all scales. We find that in the plectoneme, the short wavelength fluctuations renormalize the two moduli in a nontrivial way. On the global scale we analyze in depth the appearance of multiple plectonemes and find a sharp transition between multi-plectoneme and single plectoneme behavior. In section.~\ref{sec:comp-with-exper} we compare the model with several sets of experiments concerning extension and torque of the supercoiling molecule. We close the paper with section~\ref{sec:close} with a short discussion of some other recent models and an outlook to future developments.

\section{\label{sec:1}The Hamiltonian}
To include a twist degree of freedom, the DNA-molecule with contour length $\Lc$ is modeled as a
ribbon, or equivalently a framed space curve, defined  through its tangent $\vec{t}(s)$ and local frame rotation $\psi(s)$,  with as parameter the arc-length $s$. The number of turns $\Lk$ (linking number) we set to zero in the torsionally relaxed state. Its sign we choose positive when rotating the bead anticlockwise, as seen from the top, tightening the right handed double helix. The gradual decrease of extension of the chain under constant tension $f$, while in- or decreasing $\Lk$ from zero is well described within the framework of linear elasticity\cite{Moroz:1998}. The two moduli are the (orientational) \emph{persistence length} $\pb$ and the \emph{torsional persistence length} $\pc$. Addition of a stretch and twist-stretch coupling with a modulus that turns out to be negative\cite{Gore:2006,Daniels:2009}, slightly improves upon this. It explains why the measured extension is not fully symmetric close to the relaxed chain. We will not  consider them for the rest of the paper, since the forces below $4$ pN we deal with are small compared to the experimental stretch modulus as explained in the previous section. Anyhow addition of a coupling term does not essentially complicate the modeling. The right-handedness of the double helix on the other hand does show up in an extended plateau, already at reasonable low forces, when rotating the bead in the negative $\Lk$  direction. This is caused by the denaturation of the molecule. Since we are interested in the formation of plectonemes, we from now on restrict ourselves to the positive direction, although the same results will hold for negative $\Lk$ as long as there is no denaturation.

For the high salt-concentration, $\cs$,  persistence length we take $\pb(\infty)=\SI{50}{nm}$ to which we add the usual electrostatic stiffening corrections following OSF-theory\footnote{OSF stands for Odijk\cite{Odijk:1977} and  Skolnick, Fixman\cite{Skolnick:1977}} with a charge density along the chain limited by Manning condensation\cite{Manning:1969}:
\begin{align}
 \pb(\cs)=\pb(\infty)+\frac{1}{4\kappa^2(\cs)Q_B}
\end{align}
with $\kappa(\cs)$ the  inverse Debye screening length and $Q_B$ the Bjerrum length of the solvent:
\begin{align}
\kappa&=\sqrt{\frac{2 q_e^2 n_s}{\epsilon_r\epsilon_0 \kt}} &
 Q_B&=\frac{q^2}{4\pi\epsilon_r\epsilon_0\kt}
\end{align}
with $\epsilon_0$ the electric constant, $\epsilon_r$ the dielectric constant of water, $q_e$ the elementary charge and $n_s$ the number density of salt molecules.
For water at room temperature, the Bjerrum length is \nm{0.715}. The OSF correction is small though for example at \SI{20}{\milli\Molar} and room temperature the correction term is $\sim \nm{1.6}$.

The energy of a chain configuration up to this transition that we have to minimize has the usual elastic contributions:
\begin{align}
    \En &= \int_{-\Lc/2}^{\Lc/2} \de s \left(
        \frac{\pb}{2} \vec{\dot{t}}^2(s) + \frac{\pc}{2} \dot{\psi}^2(s) - f
        \cdot \vec{t}(s) \right) - 2 \pi \Lk([\vec{t}, \dot{\psi})]) \tau
    \label{eq:simple_energy}
\end{align}
The linking number depends on the local frame rotation around the tangent, but also on the space curve the backbone traces out when traveling along the contour. The torque $\tau$ functions here as a Lagrange multiplier. The relation between $\Lk$ and the configuration of the chain we can extract  from the celebrated C{\u{a}}lug{\u{a}}reanu-White\cite{calugareanu:1959,White:1969} relation where we imagine the chain forming part of a closed loop in which case:
\begin{align}
    \Lk([\vec{t}, \dot{\psi}]) = \Tw([\dot{ \psi}]) + \Wr([\vec{t}]).
    \label{eq:white}
\end{align}
The twist $\Tw$ is the integrated number of turns the frame rotates around the tangent direction, $\Lk$ can be written as the Gauss integral of the of the two ribbon lines while the writhe is the Gauss linking number of one of the ribbon lines with itself.:
\begin{align}
    \Tw =& \frac{1}{2 \pi} \int_0^{\Lc}\de s\dot{\psi}(s).
     &
 \Wr =& \frac{1}{4 \pi} \oint_C
   \oint_C \frac{\de\vec{r}
    \wedge \de\vec{r}'
    \cdot
    (\vec{r}-\vec{r}')}{|\vec{r}-\vec{r}'|^3}.  \label{eq:twistwrithe}
\end{align}
For the fictive loop the angle function is in general multivalued, and the writhe depends on the way we close the loop. We can overcome both problems by relying on Fuller's formula\cite{Fuller:1978,Aldinger:1995} that calculates the difference in writhe between two closed curves that are \emph{writhe homotopic}: a homotopy of curves in the space of non-intersecting closed space curves such that nowhere along the homotopy, for any given $s$, the tangent is anti-parallel is to its value at the ends of the homotopy. In that case is the writhe difference given by\cite{Fuller:1978}:
\begin{align}\label{eq:fuller}
    \Wr_2 - \Wr_1 = \frac{1}{2 \pi} \oint \de s \frac{\left(\vec{t}_1 (s) \times
    \vec{t}_2(s)\right) \cdot \left( \vec{\dot{t}_1} + \vec{\dot{t}_2}
\right)}{1 + \vec{t}_1(s) \cdot \vec{t}_2(s)}
\end{align}
This formula follows from the interpretation of the writhe as the area on the direction sphere enclosed by the tangent, when going around the loop. Fuller's formula calculates the area difference between the two homotopic curves. We consider the chain to be clamped at both ends such that the tangent and its derivative are fixed at the ends. Defining the zero of the the chain's writhe to be the torsionally relaxed state, we can calculate the writhe of any writhe homotopic perturbation under the clamped boundary conditions from Eq.~\eqref{eq:fuller} integrated over the chain, effectively keeping the closing part invariant. Implicitly we assume the bead is large enough, compared to chain fluctuations, that the chain will not change linking number by looping over the sphere.

We choose the twist angle coordinate to be zero at the substrate.
A linear stability analysis around the straight configuration is now straightforward showing there is a bifurcation point at
\begin{align}\label{eq:Lkcr}
\Lk_{\textrm{cr}}=\Lc\frac{\sqrt{f\pb}}{\pi\pc}
\end{align}
\begin{figure}[htbf] 
\centering
\subfloat[Curves of the homoclinic solutions for homoclinic parameter $t=0,1/3,2/3$ and $1$.]
  {\includegraphics[width=0.44\linewidth]{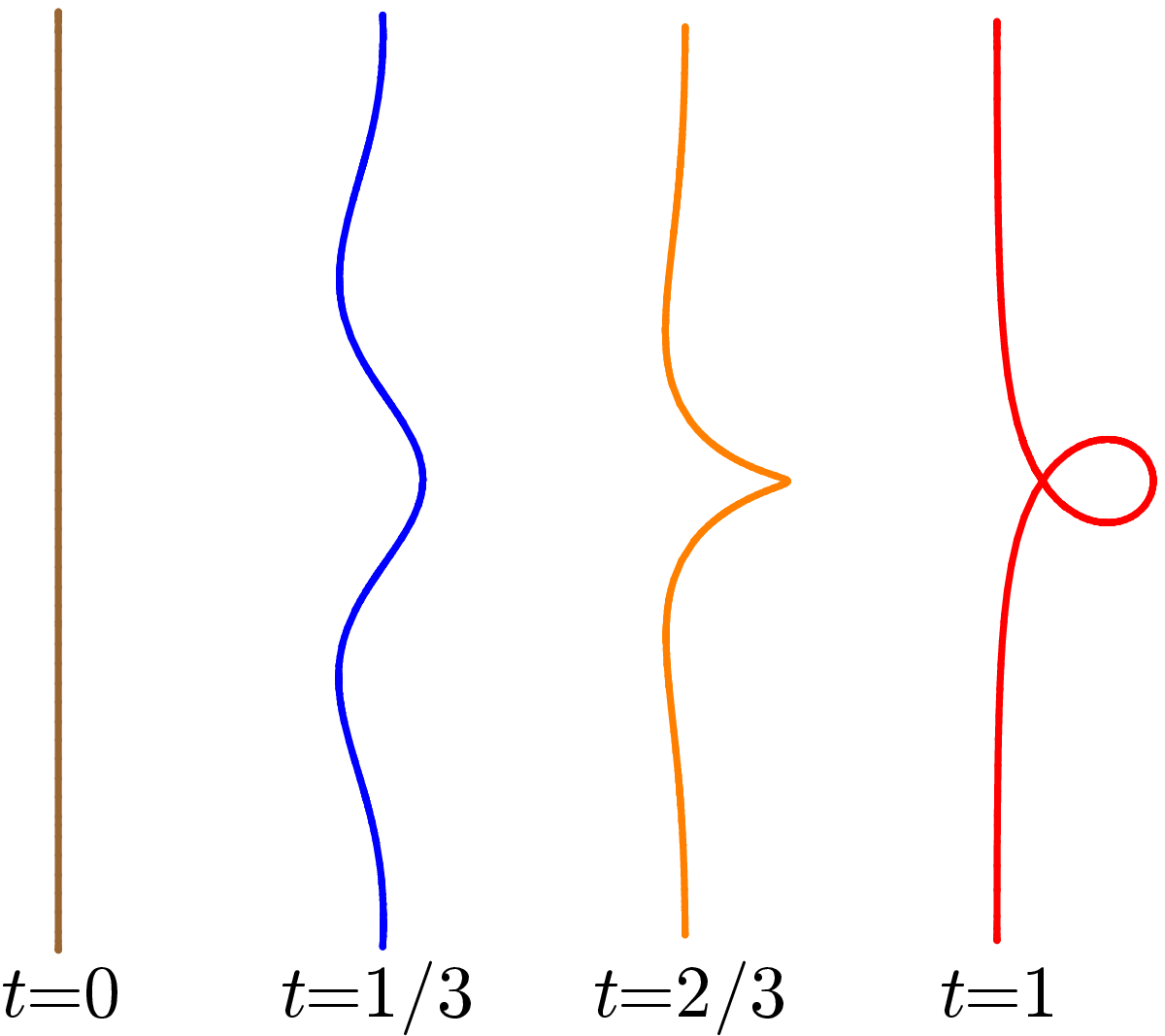} \label{fig:homoclinic}}
 \subfloat[The energy of the homoclinic solutions relative to the straight rod
  energy, where all the linking number is in the twist of the chain.  Here \(
  L_c = 600\, \)nm and $f = 2$ pN \( \gg f_0 \).]{ \includegraphics[width=0.54\linewidth]{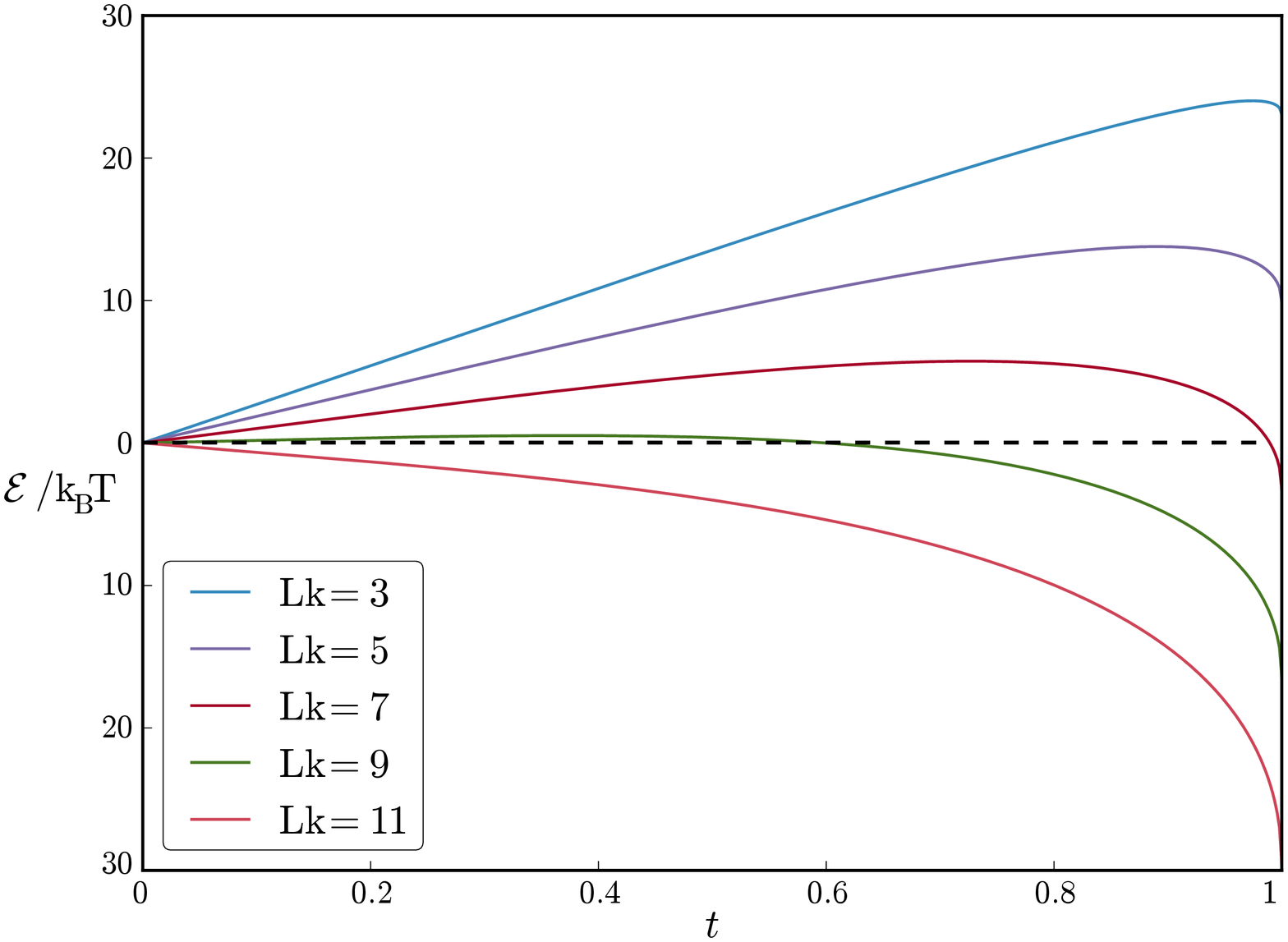}\label{fig:energy_h}}
\caption{The homoclinic solutions}
\end{figure}
Before reaching this bifurcation point other local minima start to appear, which have to be taken into account in a thermal environment. The energy minima  of the Hamiltonian that we are looking for should fulfill the boundary conditions of clamped ends with tangents parallel to the tension. The homoclinic solutions of an elastic rod under tension fulfill these boundary conditions in the infinite rod limit. They form a one parameter family of localized helices, see Fig.~\ref{fig:homoclinic}, ranging from the straight rod to a localized loop in spherical tangent coordinates given by\cite{Nizette:1999}:
\begin{align}
\begin{split}
    \cos \theta(s,t) &= 1 - 2 t^2 \sech^2 \left(\frac{s t}{\lambda}\right)  \\
    \phi(s, t) &= \arctan \left[ \frac{t}{\sqrt{1 - t^2}} \tanh\left( \frac{st}{\lambda}\right) \right] + \sqrt{1 - t^2} \frac{s}{\lambda},
\end{split}
\end{align}
with $t\in[0,1]$ and $\lambda=\sqrt{\pb/f}$ the deflection length, the length-scale above which the tension dominates  thermal fluctuations. Each of these solutions are valid for a specific torque and are not  ground-states in the supercoiling problem. They do function though as lowest  col over the barrier towards the almost closed loop that forms the start of a plectoneme.  This can be shown in a straightforward manner starting from Eq.~\eqref{eq:simple_energy}, using spherical coordinates for the tangent field. The twist term we can drop for the analysis. The writhe of the chain is a continuous map of the space curves that form the homotopy connecting the straight curve and the almost closed loop. The Euler-Lagrange equations are easy to solve using the boundary conditions $\theta(\pm\Lc/2)=0$ and solving $\tau$ for a fixed writhe, we find that the homoclinic solutions are indeed the extrema of the solutions with this writhe. The second functional derivative shows them to be minima.

When traversing the homoclinic solutions from $t=0$ to $t=1$, the writhe and bending energy of the chain are given by:
\begin{align}
\begin{split}
    \Wrl(t) &
    \simeq\frac{1}{2 \pi} \int_{-\infty}^{\infty} \de s
    \frac{\hat{e}_z \times \vec{t}(s,t) \cdot \dot{\vec{t}}(s,t)}{1 + \hat{e}_z \cdot
    \vec{t}(s,t)}= \frac{2}{\pi} \arcsin(t).
    \label{eq:wrl} \\
\Enl&=2f\Ll=8f\lambda t,
\end{split}
\end{align}
with $\Ll$ denoting the decrease in extension of the chain which we identify with the loop-length.
When keeping $\Lk$ constant the increasing writhe decreases the twist following Eq.~\eqref{eq:white}, resulting in a loop energy at constant $\Lk$ of:
\begin{align}
    \En(t) = \Enl + \frac{2 \pi^2
    \pc}{\Lc} \left( \Lk - \frac{2}{\pi} \arcsin(t) \right)^2.
    \label{eq:energy_homoclinic_lk}
\end{align}
From this expression follows that for tensions $f<f_0:=4\pc^2/(\pb \Lc^2)$ the
energy
minimum shifts from the straight rod continuously to the homoclinic loop when increasing the
linking number from $\Lk_{\textrm{cr}}$~\eqref{eq:Lkcr} till $1$. For tensions above $f_0$ only a limited range
of stable solutions in between the two extrema exists. Also in that case the
straight rod ceases to be stable at $\Lk_{cr}$, while the barrier to the loop solution
disappears a little later when:
\begin{align}
 \Lk=\Lk_{\textrm{cr}}\sqrt{1-\frac{4}{\Lk_{\textrm{cr}}^2\pi^2}}+\frac{2}{\pi}\arcsin\left(\frac{2}{\Lk_{\textrm{cr}}\pi}\right)
\end{align}
In Fig.~\ref{fig:energy_h} a typical situation is sketched for a chain of \nm{600} and a tensile force of \SI{2}{\pico\newton} $\gg f_0$. Note how already in an early
stage
a local minimum starts to form separated from the straight rod by a barrier and how that barrier moves to smaller $t$ values with increasing $\Lk$.

When $t$ approaches one, the closed loop, excluded volume interactions have to be taken into account. DNA is, at neutral pH, a strong polyelectrolyte with one charge per backbone phosphate. In a thermal environment the interaction between two chains approaching each other under a large angle is a steep potential, at a distance not far from the Debye screening length\cite{Odijk:1978}. A point of closest approach exists in homoclinic solutions
whenever $t > t_c \simeq 0.80424$. Its value, $d_{\textrm{min}}(t)$, is
the non-trivial minimum of
\begin{align}
    d(s, t) &= 2 \lambda \sqrt{4 t^2 \sech^2 \frac{s t}{\lambda} \sin^2
    \frac{s \sqrt{1 - t^2}}{\lambda} + \left( \frac{s}{\lambda} - 2 t \tanh
    \frac{s t}{\lambda} \right)^2}.
    \label{eq:d_nontrivial}
\end{align}
Within the range \( t \in [t_c, 1[ \) we can approximate $d_{\textrm{min}}$
with
\begin{align}
    d_{\textrm{min}} (t) = 2 \lambda \left( \sqrt{\frac{1 - t}{0.3799}} -
    0.00112 \right).
    \label{eq:d_min}
\end{align}
For a given force this distance has a maximum of $d_{\textrm{min}} (t_c)
\simeq 1.4 \lambda $. The point of closest approach functions a pivot point from which the plectoneme nucleates as long as it is energetically cheaper to reduce the twist through a writhing plectoneme than through the writhe of another homoclinic loop.

The radius $R$ and angle $\alpha$ of the plectoneme are set by a delicate balancing of a variety of contributions. The electrostatic repulsion, the bending energy and the entropic repulsion all depend on and influence directly $R$ and $\alpha$. Indirect they, as does the tension, influence the parameters through the writhe efficiency of the plectoneme.
This forms the basis for most of the modeling done for plectoneme formation. For the electrostatic repulsion we use the results from Ref. ~\cite{Ubbink:1999}:
\begin{align}\label{eq:plectelec}
\eel(R,\alpha)&= \frac{\neff^2Q_B}{2}
  \sqrt{\frac{\pi}{\kappa R}} e^{-2 \kappa R}
  Z\left(\cot(\alpha)\right)\\
Z(x)&= 1+m_1x^2+m_2x^4 &m_1=0.828, m_2=0.864\notag
\end{align}
valid for $\cot(\alpha)<1$, with  $\neff$ the effective charge density of the centerline of a cylinder that is the source of a Debye-H\"uckel potential that coincides asymptotically, in the 
far field, with the nonlinear Poisson-Boltzmann potential of that cylinder with
a given surface charge. For dsDNA we take a naked charge density of $2$ charges per
\nm{0.34}, representing the 2 phosphate charges per basepair, and a radius of
\nm{1}. The expansion is a fit that behaves reasonably also for $\cot{\alpha}$ close to one, where a standard asymptotic expansion would fail.
The effective charge density $\neff$ is finally calculated following Ref.~\cite{Philip:1970}.  

In contrast to the persistence length corrections, these calculations are based on the bare charge of the DNA chain. It can be shown that Manning condensation follows asymptotically\cite{Tellez:2006}.
Note  that by using an effective potential based on the Poisson-Boltzmann equation the model already includes thermal motion of counterions and salt ions. We will nonetheless refer to the model in this section as being a-thermal. 

Note further that we use the usual simplification of taking the plectoneme radius and angle to be constant
along the plectoneme. 
We set the homoclinic parameter $t$ by the
demand that the nontrivial shortest distance between the two legs of the homoclinic solution equals twice the
plectoneme radius. It is here that we will define the start of the plectoneme. The remaining part of the homoclinic solution stays connected to the end of the plectoneme and rotates around the plectoneme axis with growing plectoneme length. In this way our solution is continuous, though not in general differentiable.
One could argue that the assumption of constant plectoneme parameters does not represent the true minimum of the free energy and that in reality the space curve
should be smooth. However these are details of the energetics that are not important for the experiments, where most contributions come from the plectoneme alone.
The plectoneme has next to the potential energy density, caused by the tension, the usual energy density contributions of bending:
\begin{align}\label{eq:plectbend}
 \eb(R,\alpha)&=\frac{\pb}{2}\frac{\cos^4(\alpha)}{R^2}.
\end{align}
The writhe density of the plectoneme is given by the well known expression:
\begin{align}\label{eq:plectwrithe}
\wri&= \frac{\sin(2\alpha)}{4\pi R}.
\end{align}
This expression is often not appreciated. The naive approach of calculating the writhe density using Fullers equation relative to the plectoneme axis does result, upon averaging, in the right expression but does neglect the influence of the end loop and is relative to the wrong axes! Arguing that the end-loop is only a short stretch of the chain and thus negligible, is clearly wrong since every turn of the plectoneme length contributes equally to the writhe of the end-loop.
In the appendix~\ref{app:writhe} it is shown that in fact expression~\eqref{eq:plectwrithe} is right compared to the tension axes only \emph{when including} the end-loop contribution to the plectoneme writhe.

Putting the ingredients together we find for the energy of the chain with plectoneme:
\begin{align}\label{eq:freeplect}
  \En(R,\alpha)&= \Enl(t(R))+ \Lp\epl(R,\alpha)
+\frac{2\pi^2\pc}{\Lc}(\Lk-\Wrl-\Lp\wri)^2\\
\epl(R,\alpha)&:=f+\eb(R,\alpha)+\eel(R,\alpha).\notag
\end{align}
The plectoneme contour length $\Lp$ is found by minimizing the energy:
\begin{align}\label{eq:pllength0}
 \Lp&=\frac{\Lk-\Wrl}{\wri}-\frac{\epl \Lc}{4\pi^2\pc\wri^2},
\end{align}
where we assumed $\Lk$ to be large enough to make $\Lp$ positive.
For a long enough chain and plectoneme the loop contribution can be neglected in determining the optimal values for the plectoneme parameters $R$ and $\alpha$. This infinite chain limit is the usual approach in modeling the plectoneme and is already implicitly included in the electrostatic contribution~\eqref{eq:plectelec}. The price we pay for this simplification is small, at most noticeable close to the transition.

Starting from the torsional relaxed chain, after introducing a certain number of turns lower than the critical linking number, a solution containing a plectoneme will appear with an energy equal to the straight solution. More precisely, the buckled configuration at the transition has either a finite plectoneme that minimizes the energy or consists of only the loop:
\begin{align}\label{eq:pleclength}
\begin{split}
 L_{p,\rm{tr}}&=\begin{cases}
0& \text{in case } \Delta\leq 0 \text{ or } \sqrt{\frac{\Lc\Delta}{2\pi^2\pc}}<\Wrl\\
                 \left(\sqrt{\frac{\Lc\Delta}{2\pi^2\pc}}-\Wrl\right)\frac{1}{\wri} &\text{otherwise}
                \end{cases}\\
 \Lk_{\rm{tr}}&=\begin{cases}
                \frac{\epl}{4\pi^2\wri}\frac{\Lc}{\pc}+\sqrt{\frac{\Lc\Delta}{2\pi^2\pc}} & \text{if }L_{p,\rm{tr}}> 0\\
\frac{\Enl}{4\pi^2\Wrl}\frac{\Lc}{\pc}+\frac{1}{2}\Wrl & \text{if }L_{p,\rm{tr}}=0
               \end{cases}
\end{split}
\end{align}
with
\begin{align}\label{eq:Deltadef}
 \Delta:=\Wrl\left(\frac{\Enl}{\Wrl}-\frac{\epl}{\wri}\right),
\end{align}
the cost per writhe difference between loop and plectoneme. In this non thermal model plectoneme formation will not happen when $\Delta<0$ since it is always cheaper to form a new loop than to grow a plectoneme, but entropic contributions that we will treat in the next section will change that.
This transition point is marked by a drop in extension that is partly due to the homoclinic loop, partly due to the length of the plectoneme at the transition. Although this transition is not sharp a local minimum leading to the plectoneme does not appear until $\Lk$ has reached a value $\Lk^0$ that either:
\begin{enumerate}
 \item  the plectoneme length minimizing the energy~\eqref{eq:pllength0} has reached zero: $\Lk_0=\frac{\epl \Lc}{4\pi^2\pc\wri}+\Wrl$, or
\item  the homoclinic solution has reached the maximum of the energy barrier at $t_R$ and marking the  formation of a local minimum  at zero plectoneme-length:$\Lk^0= \Lc\frac{\pb\sqrt{1-t_R}}{\pc\pi\lambda}+\frac{2}{\pi}\arcsin(t)$.
\end{enumerate}
In any case we see that in the infinite chain limit $\Lk^0$ scales as $\Lk_{\rm{tr}}$ with the contour length. Therefore we will in the following switch to linking number densities, $\lk:=\Lk/\Lc$. 

The plectoneme length depends on both tension and salt concentration,  but on top of that scales with the square root of the contour length. This has some interesting consequences when considering the appearance of multiple plectonemes. In case the ground-state at the transition has a finite size plectoneme length, the number of  plectonemes does in general not grow with the system size. This in contrast with a situation where $\Delta\leq 0$. This will become a point size defect in the infinite chain limit and results in a finite density of plectonemes.
Roughly speaking, increasing $f$ or decreasing the salt concentration $\cs$ increases the energy per  writhe of the plectoneme, thereby shortening its start length. This leads to the following picture in the $f, \cs, \lk$ space: for high $\cs$ and low $f$ there is a first order like transition from the plectonemeless  configuration to a finite length plectoneme. The jump in extension scales with the square root of the chain length. These transition points are like a plane of first order transitions dominated by the finite length of the starting plectoneme. The plane ends in a line of continuous transitions where the transition is from straight to a configuration with an increasing number of plectonemes resulting in a finite plectoneme density: the \emph{multi-plectoneme} phase. A drop in extension caused by the end loop can still be present for short chains but thermal fluctuations smoothen the transition for longer chains.   

This ``multi-plectoneme phase'' has some interesting, biologically relevant, dynamical properties that we will come back to in the next section.

\section{\label{sec:thermal}Thermal fluctuations and the multi-plectoneme phase}

To account for thermal fluctuations several strategies have been employed in modeling  plectoneme formation. The simplest strategy is to ignore them\cite{Clauvelin:2007,Purohit:2008,Maffeo:2010} at most adding an overall  chain shortening factor\cite{Maffeo:2010} that does not change the slope. Another strategy is to ignore only thermal fluctuations in the plectoneme\cite{Neukirch:2004,Neukirch:2011}, arguing that at least for higher tensions the fluctuations are small and can as a consequence be neglected. To account for the entropic repulsion of the strands confinement entropic term from older bacterial supercoiling models is added as an independent ingredient\cite{Sheinin:2009,Neukirch:2011}.

In the first case, it is not clear why the size of the thermal fluctuations
inside the plectoneme should be the same as in the tails. The confinement of
the chain in the plectoneme is the result of a subtle equilibrium between the
applied tension, the electrostatic repulsion and the need to reduce the twist
through writhe. Furthermore this procedure needs an extra surface charge
reduction of the chain to reproduce experimental slopes\cite{Maffeo:2010}.

The second approach (fluctuations in the plectoneme are small), when properly applied, does not need this charge reduction to get a reasonable agreement with some of the experiments (as long as the salt concentration is not too low) but
has the conceptual problem that there is no a priori reason why the plectoneme
would be totally immune to fluctuations. The reasoning that thermal
fluctuations are small within the plectoneme and thus can be ignored is
erroneous since the plectoneme free energy has to be compared with the
tails where the finite fluctuations have a known
dependence on tension and applied torque.
The only conclusion one can draw, following this line of thought, is that the
extreme reduction in the number of configurations prohibits plectoneme
formation.

The last approach ignores the influence of torsion although this torsion is strongly influencing the free energy in the tails. Furthermore the bending energy density and the writhe density of the plectoneme are both affected by thermal fluctuations.
In the following we will model thermal fluctuations in the plectoneme with the
same rigor as was done previously\cite{Moroz:1998} for the tails.

\subsection{Short wave length fluctuations}
Below the transition we use the
results from Moroz and Nelson\cite{Moroz:1998}.  This  can be extended\cite{Marko:1998} with a finite stretch
modulus $S\simeq 300\,$nm$^{-1}$\cite{Sheinin:2009}  and twist stretch coupling $B\simeq-21$\cite{Sheinin:2009}. Including these moduli affects the (reciprocal) expansion parameter $K$ as introduced in Ref.~\cite{Moroz:1998}:
\begin{align}
 K&=\sqrt{f \pb-\left(\pi\pc'\lk+\frac{B f}{2S}\right)^2}.
 \label{eq:Kref}
\end{align}
with $\pc^{'} :=\pc-B^2 /S$ the effective torsional persistence length from Ref.~\cite{Marko:1998}.
The free energy density of the chain expressed in this factor can then be
written as\cite{Moroz:1998,Marko:1998}:
\begin{align}
\begin{split}
 \ftail&=2\pi^2\pc\lk^2-\frac{(f-2\pi
 B\lk)^2}{2S}-f+\frac{K }{\pb}\left(1-\frac{1}{4K}-\frac{1}{64
 K^2}\right)\\ &\simeq\ftwt-f\left(1+\frac{f-4\pi B\lk}{2S}\right) +
 \frac{1}{\lambda} \left(1-\frac{\lambda}{4\pb} -
 \frac{\lambda^2}{64\pb^2}\right),
 \end{split}
 \intertext{with the twist free energy density}
 \ftwt&\simeq2\pi^2\pc{'}\vev{\tw^2}=2\pi^2\pc^{'}\left(1-\frac{\lambda
     \pc^{'}}{4\pb^2} \right)\lk^2, \label{eq:ftwist}
\end{align}
Since the maximum tensions applied stay below $1 nm^{-1}$ $(4 pN)$ the effect of these moduli stays small thanks to the relatively strong resistance against stretching and we will drop them in the rest of the paper, by setting $S=\infty$, to decrease the clutter.

The twist energy is one of the main results of Moroz et al.\cite{Moroz:1998} who
introduced the notion of a \emph{thermally} renormalized torsional
persistence length:
\begin{align}\label{eq:effectivepc}
 \pcef(\lambda)=\left(1-\frac{\lambda\pc}{4\pb^2} \right)\pc.
\end{align}
The linking number that was put into the chain gets spread between twist and a
thermal writhe that is not symmetric around the straight twisted rod, but has a
directionality thereby decreasing the twist density apparently decreasing $\pc$. The expectation value of this thermal
writhe density, $\wdt$, and the resulting thermal shortening, $\rt$, both up
to lowest order, are given by:
\begin{align}
 \vev{\wdt}&=\frac{\pc \lambda}{4\pb^2}\lk \label{eq:tailwrithe}\\
\rt&=1-\frac{1}{2K}\biggl(1+\frac{1}{64
K^2}+\cdots\biggr)+\left(\frac{1-\coth(\frac{\Lc K}{\pb})}{2K}+\frac{\pb}{2\Lc K^2}
\right).\label{eq:moroz}
\end{align}
The last term in Eq.~\eqref{eq:moroz} is a finite size correction, that we will also drop in the following.

The validity of these expressions is limited to values of force and linking
number that make the expansion factor $K$ large enough. Moroz and Nelson
argued that for $K^2 > 3$, the error in the extension should be below $10\%$,
based on a comparison with the next term in the asymptotic expansion.

There are in fact $2$ other sources for errors: the appearance of knotted
configurations, that should have been excluded from the partition sum and
configurations with a writhe that differs a multiple of $2$ from the calculated
writhe caused by the use of Fuller's equation. For large $K$ when large
deviations from the straight rod are highly suppressed the influence of these
effects are small and we will consider a value of $K^2=3$ to be the lower
bound below which the theoretical treatment of Ref.~\cite{Moroz:1998}
breaks down.

Once a plectoneme is formed we can think of three distinct regions: the tails,
that can be treated as the straight solution, the end loop, and the plectoneme.

As shown in Ref.~\cite{Kulic:2007}, in a WLC under tension, the length of a loop,
not the contour length of the chain forming the loop, is to lowest order
unaffected by thermal fluctuations. This was shown for a loop with homoclinic
parameter $t=1$ with the two tails bound by a gliding ring at the contact point.
There is no reason to doubt that this will
hold also for the end loops of the plectonemes, since they are sufficiently
close to the closed loop, with the essential difference that the tails are not
bound together but lie in an effective potential well resulting from a twist
induced attraction and an electrostatic repulsion.  Thermal fluctuations
necessarily open the loop from its ground state value, thus decreasing its
length. This loop destabilization effect becomes unimportant for a finite size
plectoneme configuration, since loop opening and plectoneme radius are linked.
To avoid unnecessary complications we will just ignore the entropic loop
contributions and instead determine the relevant loop size from the plectoneme.
It is possible to add  electrostatic interactions to the loop\cite{Cherstvy:2011}, but the advantage of not having to estimate these and entropic repulsion to the end-loop free energy more than compensates for the small
error it might produce in the free energy close to a possible plectonemeless
loop configuration. In general this simplification hardly affects the jump in
length seen in the turn extension plots at the transition, since jumps indicate
usually a finite size plectoneme at the transition, while the plectoneme parameter has only a limited range in light of the lower limit $t_c$.

The plectoneme part needs a more careful examination. We start from the
calculations from Ref.~\cite{Ubbink:1999}. They considered one strand
of the regular plectoneme fluctuating in the mean field potential of the
opposing strand, assuming the fluctuations to have a Gaussian distribution
around their average in two directions perpendicular to the strand. One
direction is chosen pointing towards the opposing strand, the radial direction,
the other normal to this direction, the pitch direction.  Fluctuations in the
radial direction are dominated by the exponent of the electrostatic
interactions, while fluctuations in the pitch direction have much less
influence on the energetics. We stress the advantage of this approach over the
expansion of the effective confining potential around the ground state. In the
radial direction the potential is highly skewed, exponentially increasing
towards smaller radius.  A harmonic approximation would only be valid in a tiny
region around the ground state.  Instead we assume fluctuations small compared
to its typical length-scale, the persistence length.  Denoting the standard
deviation of the Gaussian distribution in the radial and pitch direction by
respectively $\sigma_r$ and $\sigma_p$, the electrostatic part of the free
energy changes approximately to\cite{Ubbink:1999}:
\begin{align}\label{eq:plectelecentr}
   \fel(t,\alpha,\sigma_r)&=\eel e^{4\kappa^2\sigma_r^2}= \frac{\neff^2Q_B}{2}
   \sqrt{\frac{\pi}{\kappa R(t)}} e^{4\kappa^2\sigma_r^2-2 \kappa R(t)}
   Z\left(\cot(\alpha)\right).
\end{align}
The steep exponential rise of this free energy contribution clearly
limits the value of $\sigma_r$ to be of order $(2\kappa)^{-1}$. This
distinguishes the magnitude of radial fluctuations from those in the pitch
direction.

It was argued\cite{Odijk:1998} that the standard deviation in the pitch
direction should be of the order of the pitch itself. This result one expects also on geometrical ground, as shown in Fig.~\ref{fig:plectoneme}. While an exact value is hard to obtain, it is
considerably larger than $ \sigma_r$. As it is the tightest direction that
dominates the free energy of confinement\cite{Emanuel:2013}, our results are fairly
insensitive to its precise value. In the following we chose $\sigma_p=\pi R\sin(\alpha)$,
which is the standard deviation of the channel formed by the two neighboring stretches of fluctuating opposing strand.
The undulating chain contracts with a factor $\rp$, that we will discuss
further below. This contraction on the other hand decreases the bending energy density and the writhe
density of the plectoneme in a nontrivial way.
In appendix~\ref{ap:fluctuations} it is shown that they change to:
\begin{align}\label{eq:effbendwrithe}
 \eb&\rightarrow\fb=\rp^4\eb=\rp^4\frac{\pb}{2}\frac{\cos^4(\alpha)}{R^2} &
\wri&\rightarrow\rp\wri=\rp\frac{\sin (2\alpha)}{4\pi R}.
\end{align}

To compute the entropic cost of confinement, we cannot neglect the twist in the
chain.
The twist along the backbone  couples to the other degrees of freedom mostly through the global constraint encoded in White's equation~\eqref{eq:white}. As one expects,  and was experimentally shown\cite{Crut:2007}, twist relaxation is fast compared to tangential fluctuations.  This allows us to integrate out these fast modes and take the twist free energy density to be constant throughout the chain.

In the tails thermal writhe is suppressed by the tension, while in the plectoneme it is suppressed by the confinement caused by a combination of electrostatics, tension bending and twist. These thermal writhes are in general not the same even when their twist energy densities are. Therefore we need to take the thermal writhe in the
plectoneme explicitly into consideration. We assign part of the total linking
number to the tails and loop, from which follows a tension dependent
expectation value of thermal writhe and twist density according
to Eq.~\eqref{eq:tailwrithe}.  The rest of the linking number has to be
accounted for by the plectoneme. We use this difference as the definition of
its linking number. For a large part this linking number is stored in the twist and
writhe of the zero temperature plectoneme, but partly it sits in the thermal
writhe of the strands of the plectoneme. For the calculation of the relevant
quantities of a torsionally constrained confined WLC we assume we can capture
the physics of confinement of the plectoneme strands with that of a chain
confined by a harmonic potential with the same standard deviations $\sigma_r$
and $\sigma_p$. In other words: the transversal distribution is Gaussian
enough. The relevant calculations for the confinement problem were performed in Ref.~\cite{Emanuel:2013}. The free
energy density of a confined WLC as function of linking number density $\lks$ and the standard deviations in two orthogonal channel directions
$\sigma_r$ and $\sigma_p$ is to lowest order:
\begin{align}\label{eq:freestrand}
\fstr&=\ftwp+\frac{3}{8}\left(\frac{1}{\lambda_r}+\frac{1}{\lambda_p}
\right) \\
\text{with }\qquad\ftwp&:=2\pi^2\pc\vev{\tws^2}=
2\pi^2\pcef(\lambda_s(\sigma_r,\sigma_p))\lks^2,\notag
\end{align}
where $\pcef()$ is the same function of $\lambda$ as given by
Eq.~\eqref{eq:effectivepc}. The effective deflection length $\lambda_s$, the
length scale over which the confining potential starts to dominate thermal
fluctuations, is given by\cite{Emanuel:2013}:
\begin{align}\label{eq:torsiondeflection}
\lambda_s&=2\frac{\lambda_r^3\lambda_p + \lambda_r^2\lambda_p^2 +
\lambda_r\lambda_p^3}{(\lambda_r+\lambda_p)(\lambda_r^2+\lambda_p^2)}
&\lambda_{r,p}& := ( \pb\sigma_{r,p}^2)^{1/3}.
\end{align}
The first term of Eq.~\eqref{eq:freestrand} is the twist free energy density,
the second term is the entropic cost of confinement. Note that the confining
potential, due to bending and electrostatics, is \emph{not} included\cite{Ubbink:1999,Emanuel:2013}.  To the
same order, the contraction of the polymer is found to be
\begin{align}
\rp&=1- \frac{1}{4}\left[\frac{\lambda_r}{\pb} +
    \frac{\lambda_p}{\pb}\right].
\end{align}
which is up to this order equal to the torsion-less contraction; inclusion of stretch and stretch-twist moduli or higher order terms changes this.
From Eq.~\eqref{eq:torsiondeflection} we see that in case $\sigma_r\ll\sigma_p$ the effective
deflection length reduces to $\lambda_s\simeq 2\lambda_r$ and indeed it is the
tightest direction that sets the free energy as alluded before.

These results are valid for undulations in, and thermal writhe with respect to,
a straight channel. However the writhe, as a local observable, is only defined
with respect to a reference curve, which is the writhing plectoneme. In
appendix~\ref{ap:fluctuations} it is shown that, under reasonable
assumptions, thermal writhe can be treated as an additive correction to the
plectoneme writhe, where the thermal writhe is calculated as the thermal writhe
of an undulating chain with a finite linking number, confined to a straight
channel.

The reason is that the length scale over which the fluctuation channel axis can be considered straight is of the order of the contour length over which  the $r$ and $p$ directions
rotate around the channel axis which is of the order of the pitch or, as argued above, the standard
deviation in the pitch direction. In all relevant cases is the standard
deviation in the radial direction considerably smaller than in the pitch
direction.  Since it is this length scale, associated to the tightest
direction, that determines the influence of confinement on the free energy, the
energetics of the global writhing path decouples from the thermal fluctuations.
The contraction $\rp$ depends on fluctuations in the pitch direction and
therefore its size does affect plectoneme formation. The free energy density of
the plectoneme is the sum of this confinement, the
bending~\eqref{eq:effbendwrithe} and electrostatic%
~\eqref{eq:plectelecentr}
free energy:
\begin{align}
 \fpl=\fb+\fstr+\fel.
\end{align}
Equating $\ftwt$ and
$\ftwp$ allows us to eliminate the linking density of the strands in the
plectoneme as parameter and write $\lks=(1-\epsilon)\lk$, with
\begin{align}
\epsilon&=1-\sqrt{\frac{\pcef(\lambda)}{\pcef(\lambda_s)}}
\end{align}
small but in general nonzero. This is indeed the case in all
experimental conditions studied: For forces ranging from
\SIrange{0.5}{4}{\pico\newton} and salt concentrations from \SIrange{20}{320}{\milli\Molar}, a crude estimate is easily made, namely $\epsilon \in [0,0.1]$. Although the
difference in `thermal waste' while transforming linking number into twist is
rather small, it would be wrong to draw the conclusion that entropic effects
can be neglected, since the entropic part of the free energy goes as $\simeq
\kt/\lambda$. The difference between the two states can be up to one $\kt$ per
\si{\nm}.

It is worthwhile to split off the twist contribution to the free energy densities:
\begin{subequations}
\begin{align}
\left. \begin{array}{c}
\ftail  \\
\fpl
 \end{array}\right\rbrace &= \ftw + \left\lbrace\begin{array}{c}
\gtail  \\
\gpl
 \end{array}\right.
\end{align}
with:
\begin{align}
\begin{split}
 \gtail&=-f+\frac{1}{\lambda}\left(1-\frac{\lambda}{4\pb}-\frac{
\lambda^2}{64\pb^2}\right)\\
\gpl&=\frac{3}{8}\left(\frac{1}{\lambda_r}+\frac{1}{\lambda_p}
\right)+\fb+\fel
\end{split}
\end{align}
\end{subequations}
the remaining free energy contributions. We will use $\dg=\gpl-\gtail$ to
denote their difference.  Once a plectoneme has formed the expectation value of
its contour length follows from the combined linking numbers of plectoneme and
end-loop, which should add to the linking number that was externally applied:
\begin{align}\label{eq:plectlength}
 \Lk&=(\Lc-\Lp)\lk+\Lp[\rp\wri+(1-\epsilon)\lk]+\Wrl \Rightarrow&
\lp:=\frac{\Lp}{\Lc}=\frac{\nu-\lk-\Wrl/\Lc}{\rp\wri-\epsilon\lk},
\end{align}
with $\nu:=\Lk/\Lc$ the applied linking number density.  The reduced free
energy density of this one plectoneme configuration and its extension are:
\begin{align}\label{eq:simpleext}
\begin{split}
 \free_1&= (1-\lp)\ftail+\lp\fpl\\
&= \ftw+\gtail+\lp\dg+\frac{\Enl(t)}{ \Lc}
\end{split}\\
\bar{z}&:=\frac{z}{\Lc}=\rt(1-\lp)-\frac{\Ll}{\Lc}
\end{align}
both depending on the $4$ parameters $R$ (or $t$), $\sigma_r$, $\alpha$ and
$\lk$.  The calculation boils down to a $4$ parameter minimization procedure.
\begin{figure*}[htb]
  \centering
\includegraphics[width=\textwidth]{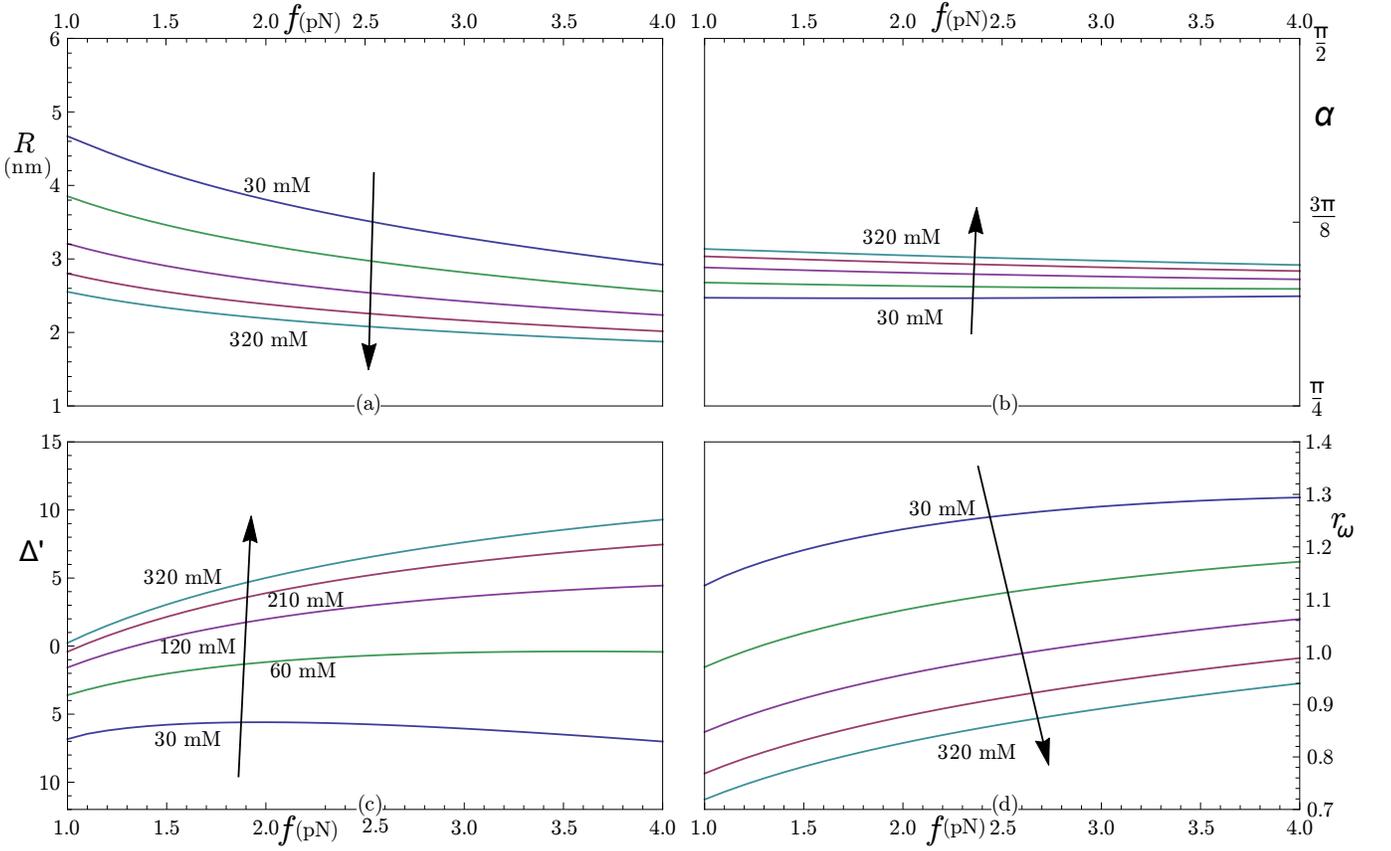}
  \caption{Force dependence of (a) the plectoneme radius, (b) the plectoneme angle, (c) the energy per writhe difference between loop and plectoneme with logarithmic correction related to the choice of cutoff, and (d) the writhe density ratio between loop and plectoneme for salt concentrations of $30,60,120,210$ and $320$ mM. The arrows point in the direction of increasing salt. The range for $\alpha$ was on purpose chosen to be the full allowed range for a stable plectoneme, showing that its value is hardly dependent on the environment}\label{fig:parameters}
\end{figure*}
The resulting plectoneme angle is almost independent of applied tension or salt concentration, see Fig.~\ref{fig:parameters}(b). This is a result of the $Z(\alpha)$ term in Eq.~\eqref{eq:plectelec} reflecting the influence of the electrostatic repulsion to counter the demand for writhe efficiency (low $\alpha$). Using this concept of energy per writhe gained in the plectoneme also helps in understanding the general trend of the plectoneme radius as shown in Fig.~\ref{fig:parameters}~(a). Increasing the tension decreases the radius to counter the growing energy per writhe. The same holds for an increase of the range of the electrostatic repulsion, by lowering the salt concentration.
Note that the plectoneme radius is always large enough for the (reduced) electrostatic potential to be below one in the overlap region in between the two strands. This is needed to justify the use of the Debeye-H\"uckel tails in calculating the effective potential between the strands\cite{Hoskin:1956}.

In the long chain limit with finite plectoneme length the loop contribution can
be neglected in determining the $4$ parameters. We can assume that $\epsilon$
is small compared to $\wri$, under conditions where a plectoneme forms.
We can also neglect the dependence of $\rp$ on the parameters, its variational
contribution is on the order of $\lambda_{r,p}/\pb$, which is small by
assumption.  The long chain finite plectoneme free energy is:
\begin{align}\label{eq:freethermal}
\free_1&=\ftw(\lk) +\gtail
+\frac{\nu-\lk}{\rp \wri(R,\alpha)} \dg(R,\alpha,\sigma_r)
\end{align}
The linking number density and chain extension are readily obtained in this
limit:
\begin{align}\label{eq:lkthermal}
 \lk &= \frac{\dg}{4\pi^2 \pc\rp\wri(R,\alpha)} & \bar{z}&=
 -\frac{\rt}{\rp\wri(R,\alpha)}
\end{align}
Minimizing the free energy is within this approximation equivalent to
minimizing the linking number density. This is not really a surprise since
plectoneme formation is driven by linking number.

A numerical minimization gives
results that compare reasonably well with experiments.  The transition point, height of the jump at the transition
as well as the slope after the transition are within experimental error for
high enough forces and salt concentrations, see the dotted lines in
Fig.~\ref{fig:totslope}.  The lack of agreement at low salt clearly inversely correlates
with $K^2$. Dropping the assumption of equal linking number densities in tail
and plectoneme hardly improves the results, even when the value of $K^2$ stays well above $3$. This discrepancy, that is slightly stronger when fluctuations are neglected, has led to a variety
of speculations, like an effective charge reduction\cite{Maffeo:2010}, or a charge correlation effect
between the two intertwined super-helices that form the
plectoneme\cite{Neukirch:2011}. The deviation of the experimental slopes from
the calculated one goes hand in hand with the decrease of the height of the
potential barrier between straight and plectoneme configuration.  But our
theory is not complete yet: the inclusion of other local minima next to these
two configurations turns out to be of greater importance than has been
acknowledged until now, as we will show in the next section.

\subsection{Tunneling to the plectoneme}
Contributions of local minima have to be taken separately into account in any perturbative calculation.  Accepting the simplification that a plectoneme has a well defined radius and angle that are length independent, the only concern is the barrier height between $t=0$ and its final value $t_R$ corresponding to the plectoneme radius.

The usual way to take these local minima into account is to treat them as a gas
of defects that compete with their entropic gain against the energetic
advantage of the ground state. This is the situation that would exist in a
torque regulated setup. In our case where the linking number is the control
parameter the treatment changes essentially. A defect changes the linking
number and so the energy of the configuration in which it is embedded.
Furthermore the defects are themselves plectonemes and
so to understand thermal fluctuations close to the transition we actually
study multi-plectoneme configurations. Multiple plectonemes were considered before\cite{Daniels:2009,marko:2012} but mostly seen as small corrections on the one plectoneme configurations.

The entropic gain of a multi-plectoneme configuration is twofold: there is the
usual combinatoric positional freedom of defect placement (the ``gas of
defects''), but there is also an increase in configurations due to the freedom
in distributing the total plectoneme length over the individual plectonemes.
Treating the plectonemes as having a hardcore repulsion, one finds for the
partition sum of a configuration with \emph{total} plectoneme contour
length $\Lp(m)$ spread out over $m$ plectonemes:
\begin{align}\label{eq:hac}
\begin{split}
 Z_m&=\frac{\sqrt{\Lc}}{\Lambda^{2m-3/2}} \frac{\Lp^{m-1}(m)}{(m-1)!}
 \frac{(\Lc-m\Ll-\Lp(m))^m}{m!}e^{-\Lc\free_m}\\
\free_m&= \ftw +\gtail+\lp(m)\dg+m\Enl\\
\end{split}
\end{align}
with $\Lambda$ a cutoff scale for which we we choose the helical repeat, as explained in appendix~\ref{ap:multi} where the above expression is derived.

To streamline the notation we define the following densities:
\begin{subequations}
\begin{align}
&\text{the relative linking density:}&  r_{\nu}&:=\frac{\nu-\lk}{\rp\wri-\epsilon\lk}\\
&\text{the relative writhe density:}& r_{\wri}&:=\frac{\wrl}{\rp\wri-\epsilon\lk}\\
&\text{the loop writhe density:}& \wrl&:=\frac{\Wrl}{\Ll}\\
&\text{the loop density:}& \mu&:=\frac{m \Ll}{\Lc}
\end{align}
\end{subequations}
The $m$ dependent plectoneme length follows as before
from the total linking number:
\begin{align}\label{eq:Lpmult}
 \lp(m)=\frac{\nu-\lk-m \Wrl/\Lc}{\rp\wri-\epsilon\lk}=  r_{\nu}-r_{\omega}\mu,
\end{align}
We cannot drop the loop contribution here since we should leave the possibility
open that the number of plectonemes increases at the same (or higher) rate as
the contour length, reaching some finite density. For the same reason we  also keep the end-loop energy.  In principle also plectonemes with a negative writhe plectonemes should
be included, but their contribution is very small and practically only present when tension and linking
number are low. We are mainly interested in linking numbers around and above
the bifurcation point, thus we can neglect them.

The maximum number of plectonemes can never be higher than $\Lc/\Ll$ and
it is to be expected that finite size effects easily dominate the turn extension
curves for shorter chains. We want to describe the generic behavior of the turn
extension plot without end effects. The reason is not only to avoid
plectoneme-plectoneme interactions, but also to avoid interactions of the
magnetic/optical bead with the substrate and details of the exact geometry of
attachment of the chain ends. We write the free energy of the
chain as:
\begin{align}
\Free&=\Lc \free_0 +m\Delta=\Lc(\free_0
+\frac{\mu}{\Ll}\Delta),
\end{align}
with $\Delta$ as in Eq.~\eqref{eq:Deltadef} and $\free_0$ collecting the terms of the free energy density, that do not
depend on the loop density.

\begin{figure}[htb]
\centering
\subfloat[Contour plot of the multi-plectoneme factor as it depends on salt concentration and tension. The thick red line can be interpreted as the border between single plectoneme phase on the right and a multi-plectoneme phase on the left. The inset shows $\zeta$ as a function of salt concentration for three different forces. Note the crossing of the lines at low salt concentrations.]{
\includegraphics[width=0.45\linewidth]{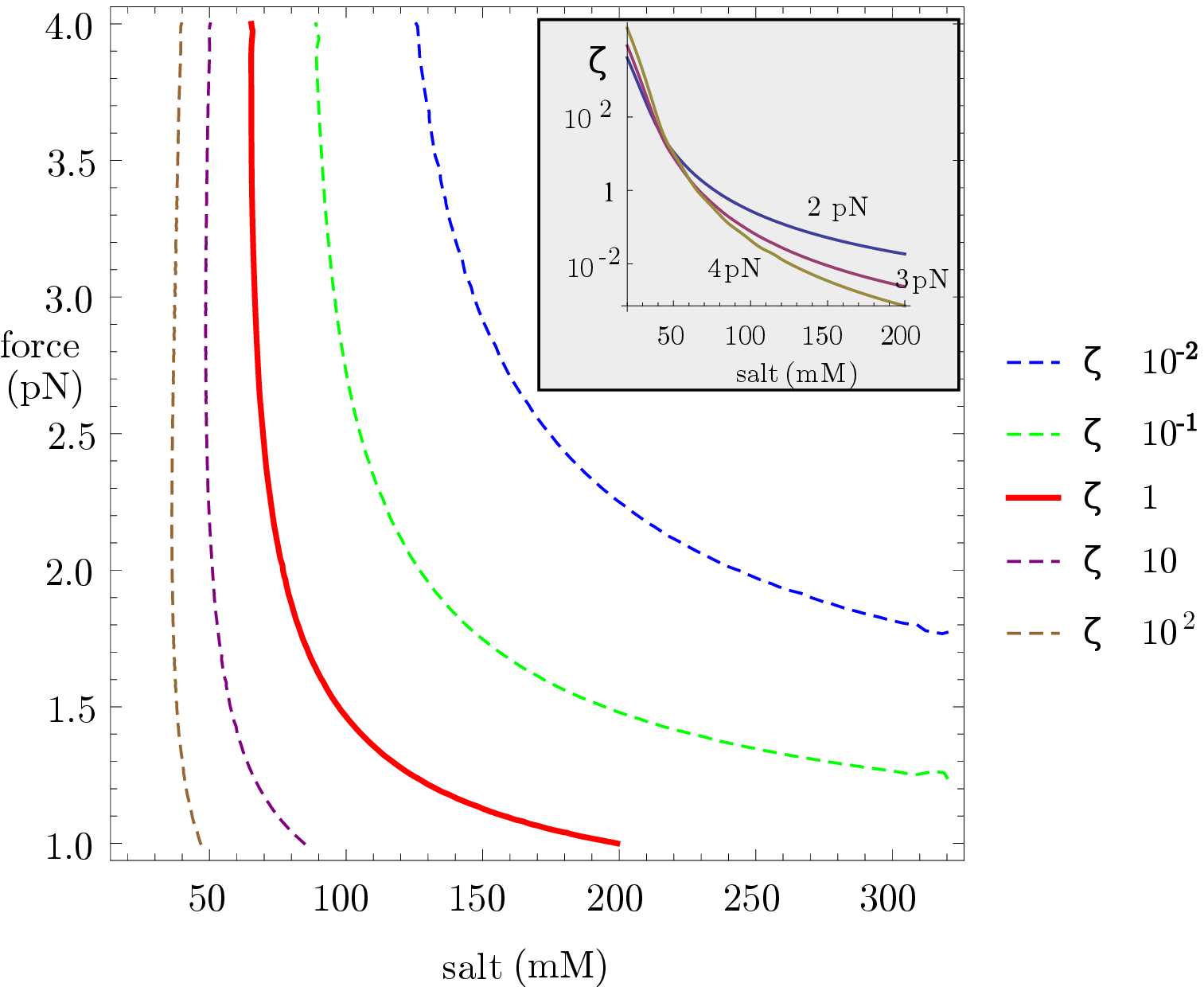}
\label{fig:mpfac}}
\subfloat[The maximal loop density $\mu$. Note the sharp transition from a maximal possible ($\mu=1$) to a vanishing number of plectonemes. Since the maximal $\mu$ is reached at the end of the plectoneme slope for $r_{\wri}>1$, it does not reflect the multi-plectoneme transition along the beginning of the slope.]{
    \includegraphics[width=0.5\linewidth]{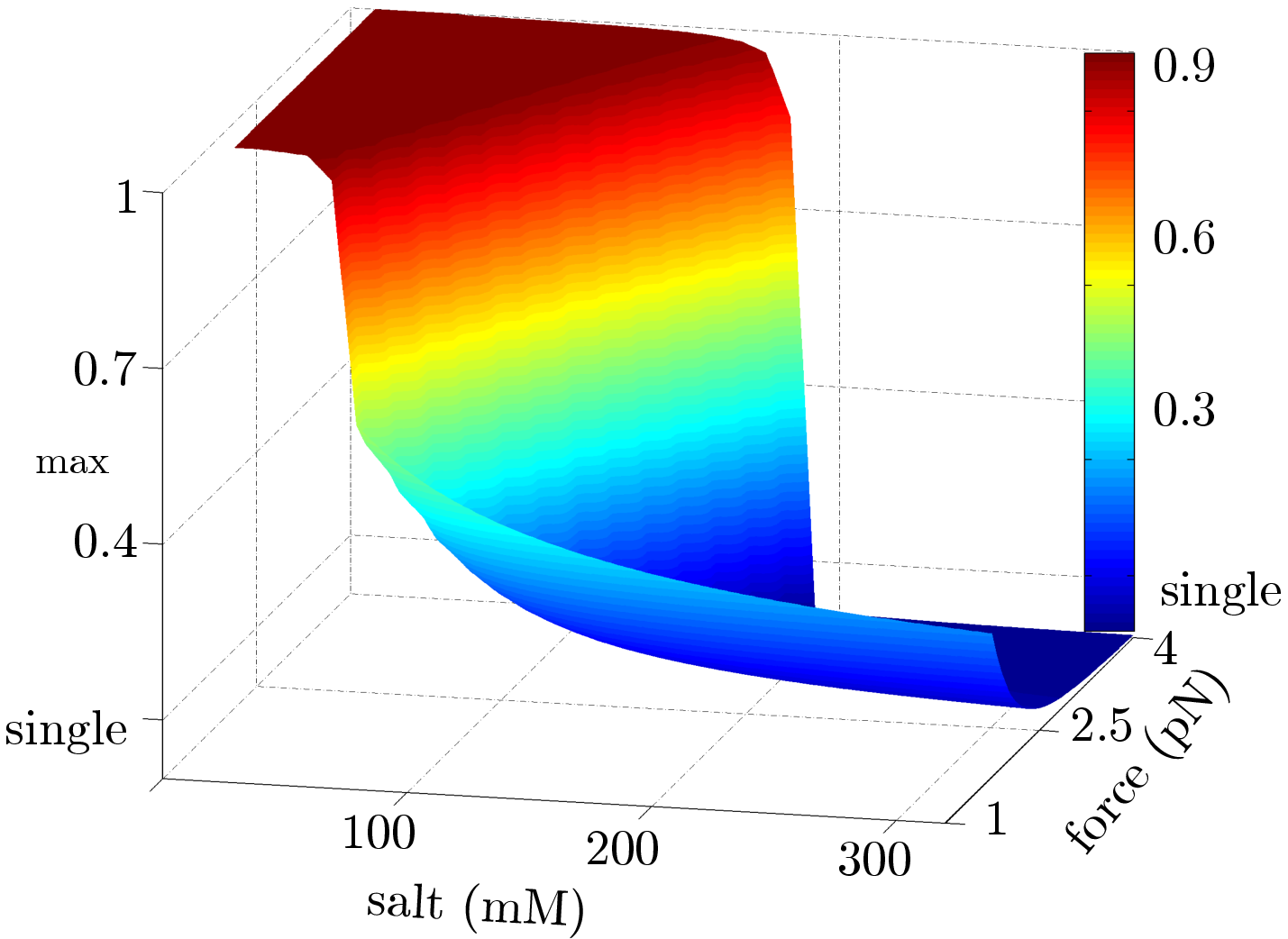}
\label{fig:mu}}
\caption{The two faces of multiple plectonemes}\label{fig:twoface}
\end{figure}

Assume we are far enough in the plectoneme region that only terms with $m>1$ contribute. The loop density dependence of the total partition sum reduces to:
\begin{align}
 Z&\sim \int_0^{\mu_m}\de\mu \exp\left\lbrace
     \frac{\Lc\mu}{\Ll}\left[
         \log\left(\frac{\lp(\mu)z(\mu)}{\mu^2}\right)+2\log(\Ll/\Lambda)+2-\Delta\right]\right\rbrace,
\end{align}
with $\mu_m$ the maximum density set by
$\mu_m=\text{sup}\lbrace\mu\in[0,1]\vert 0\leqq \lp(\mu) \leqq 1-\mu\rbrace$.
It is straight forward to verify that the argument of the exponent is
a concave function of $\mu$ is for $\mu\in(0,\mu_m)$ and so its dominant
contribution comes from its maximum:
\begin{align}\label{eq:mextr}
 &\log\left(\frac{l_p(r_{\nu},\mu) \bar{z}(r_{\nu},\mu)}{\mu^2}\right) -
 \mu\left(\frac{r_{\wri}}{l_p(r_{\nu},\mu)}+\frac{1-r_{\wri}}{\bar{z}(r_{\nu},\mu)}\right)-\Delta'=0& \Delta'&:=\Delta-2\log\left(\frac{\Ll}{\Lambda}\right)
\end{align}
Since the relative extension of the chain is $\bar{z}=1-r_{\nu}-(1-r_{\wri})\mu$ it follows that in case $r_{\wri}=1$ the turns extension slope does not depend on the number density of plectonemes. Based on our model the value of $r_{\wri}$ is often close to one (Fig.~\ref{fig:parameters} (d)). This is one reason why the appearance of multiple plectonemes took so long to discover. The energy per writhe can at the same time differ considerably between loop and plectoneme ($\Delta\neq 0$ Fig.~\ref{fig:parameters} (c)) changing the torque after the transition even when the slope can be fitted with just one plectoneme.
The more detailed analysis of Eq.~\eqref{eq:mextr} is left for appendix~\ref{app:Lambert}.
\begin{figure}[htb]
  \centering
\includegraphics[width=\textwidth]{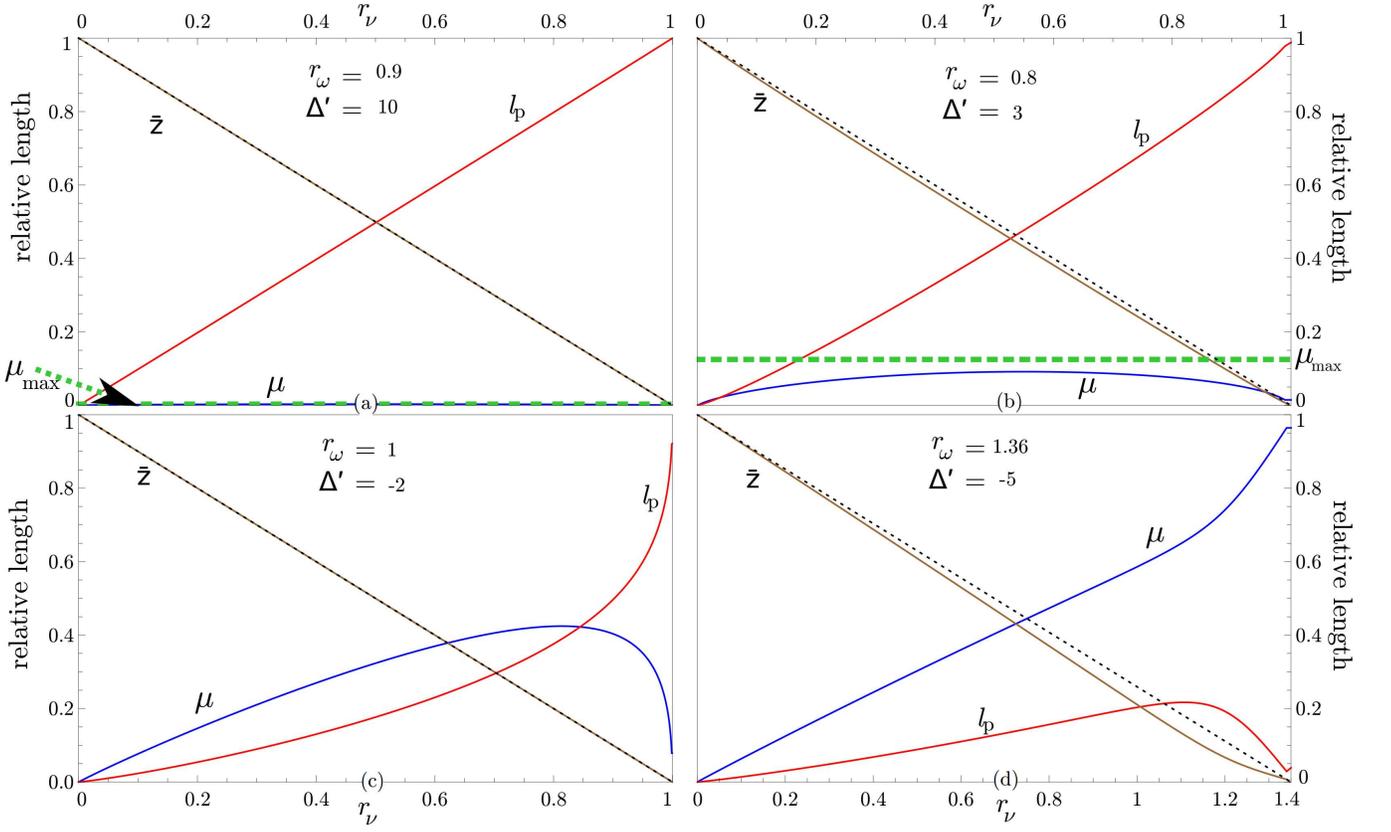}
\caption{The decrease of the extension, $\bar{z}$,  and the contributions in which it decomposes ($\mu$ and $lp$) as function of scaled linking number density, omitting the straight solution that disappears early on. The plots were generated using the parametrization outlined in appendix~\ref{app:Lambert}. The green dashed line is the approximation from Eq.~\eqref{eq:mumax}. The dotted line corresponds to the fictive one plectoneme behavior. (a) Typical single plectoneme behavior at high cost per writhe difference between loop and plectoneme. (b) Lower $\Delta$ increases the number of plectonemes, changing the slope of the turns extension plot. (c) When the ratio of the writhe densities is one the slope does not change even with a large number of plectonemes. (d) When $r_{\wri}$ rises above one we end up with a high density of zero length plectonemes. The force extension curve resembles also here a one plectoneme curve but one with modified plectoneme parameters.}\label{fig:multiplot}
\end{figure}
Some examples of the dependence of $\lp,\mu$ and $\bar{z}$ as function of $r_{\nu}$ for several combinations of $\Delta'$ and $r_{\wri}$ are shown in Fig.~\ref{fig:multiplot}.
The values of $\Delta'$ and $r_{\wri}$ corresponding to typical experimental conditions can be read of from Fig.~\ref{fig:parameters} (c) and (d). It is clear that lowering the salt concentration drives the two strands further apart thereby decreasing the energetic cost efficiency for writhe production of the plectoneme and even becoming more costly than the loop itself for low salt conditions. The formation of plectonemes in that case can be seen as a purely entropic effect. The influence of the tension is a bit more subtle. The tension increases the loop energy $\sim\sqrt{f}$, while in the plectoneme the behavior depends on the salt concentration. At high salt it is only the potential (force) term that changes, since there is not much room for changing the radius, while at low salt $R$ has more possibilities to adapt, increasing the electrostatic and bending contributions as well with increasing tension. This is only partly compensated for by an increasing writhe density in the plectoneme.
One result of practical use is the multi-plectoneme factor $\zeta$:
\begin{align}
 \zeta:=r_{\wri}^2e^{-\Delta'}.
\end{align}
As shown in appendix~\ref{app:Lambert} it functions as an indicator for the growth of multiple plectonemes soon after the transition. When $\zeta=1$ can be interpreted as the boundary between single plectoneme and multi-plectoneme behavior. Its salt and tension dependence is depicted in Fig.~\ref{fig:mpfac}.  The largest factor is at low salt and high tension, while
for high salt concentrations, $\zeta$ \emph{increases} with decreasing tension. A simpler quantity is the maximal number density for a given tension and salt concentration. Its behavior is depicted in Fig.~\ref{fig:mu}. Its change from single plectoneme to multi-plectoneme is also very sharp, but part of this multi-plectoneme behavior happens only at the end of the extension - number of turns plot, when $r_{\wri}$ is larger than one.

Towards the end of the slope the number density $\mu$ goes to zero for $r_{\wri}$ smaller or equal to one while for $r_{\wri}>1$ the plectoneme length goes to zero due to the increasing number of plectonemes. This is of course a result of the disappearing of any entropic gain when all of the chain participates in supercoiling. It is interesting to observe that the slopes can for practically all measurements be (often falsely) interpreted as a single plectoneme slope. In case the writhe ratio is above one, the plectoneme parameters have to be changed for example by modifying the electrostatic repulsion.

\subsection{Dynamics }
At the transition there are two states with equal energy that differ in extension and are separated by an energy barrier. The inclusion of an explicit and realistic loop model allows for an estimate of the transition time from one configuration to the other. Since as argued before the minimal energy path from the straight chain to the plectoneme runs over the family of homoclinic solutions with their free energies given by Eq.~\eqref{eq:energy_homoclinic_lk}, we can use the one-dimensional Kramers' equation\cite{Risken}, with some adjustment for the non-analytic potential around the straight configuration, to calculate the transition time between the two states. Although the diffusion coefficients needed to calculate the attempt frequencies are not a priori clear, the force dependence can be inferred. The transition times for DNA are at the moment too fast to extract them from available measurements, but with new measurements on the way we plan to come back to this issue in the near future.

\begin{figure}[htp]
 \centering
\subfloat[]{
 \includegraphics[width=0.53\linewidth]{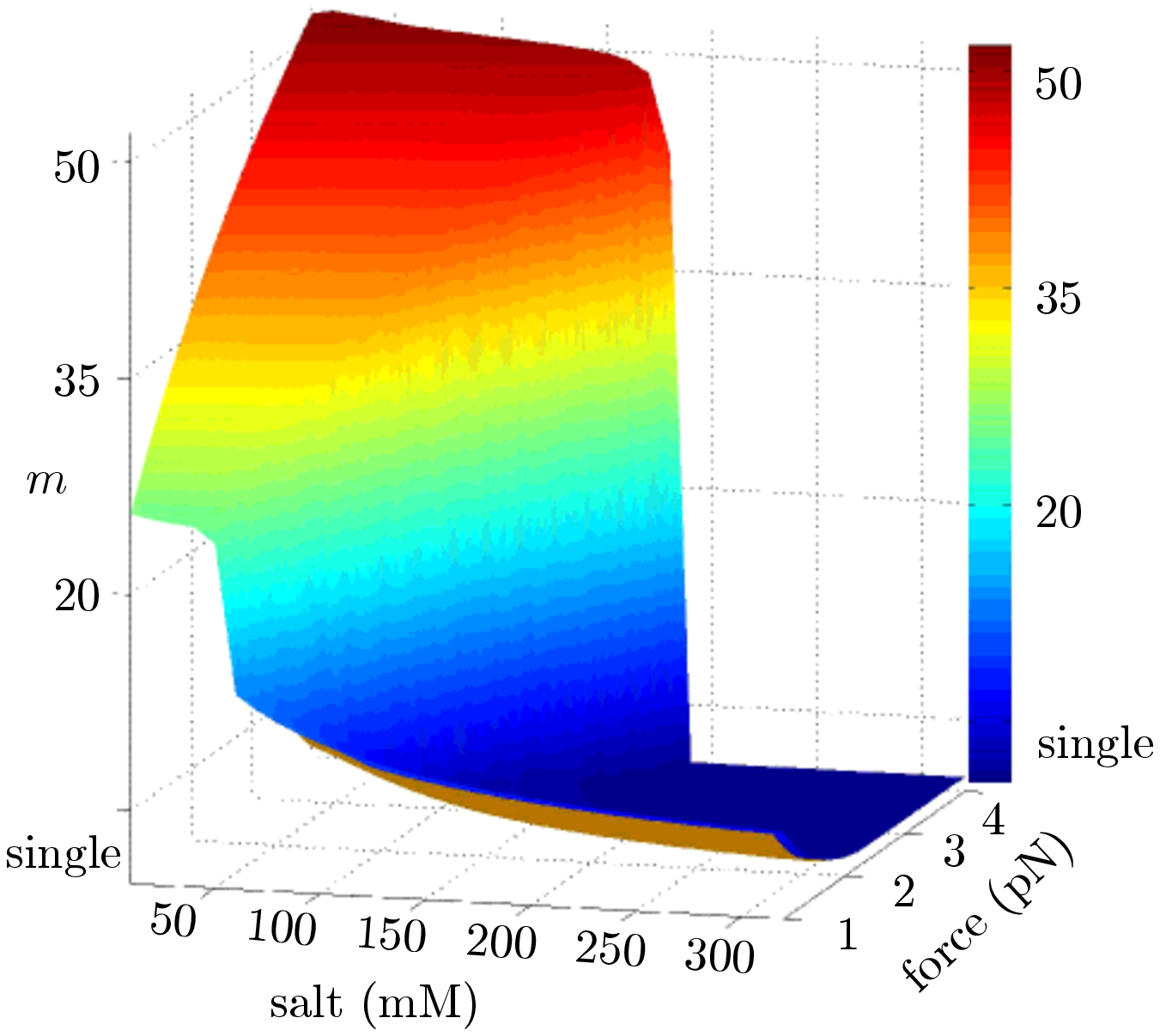}\label{fig:m1400}}
 \subfloat[]{
 \includegraphics[width=0.45\linewidth]{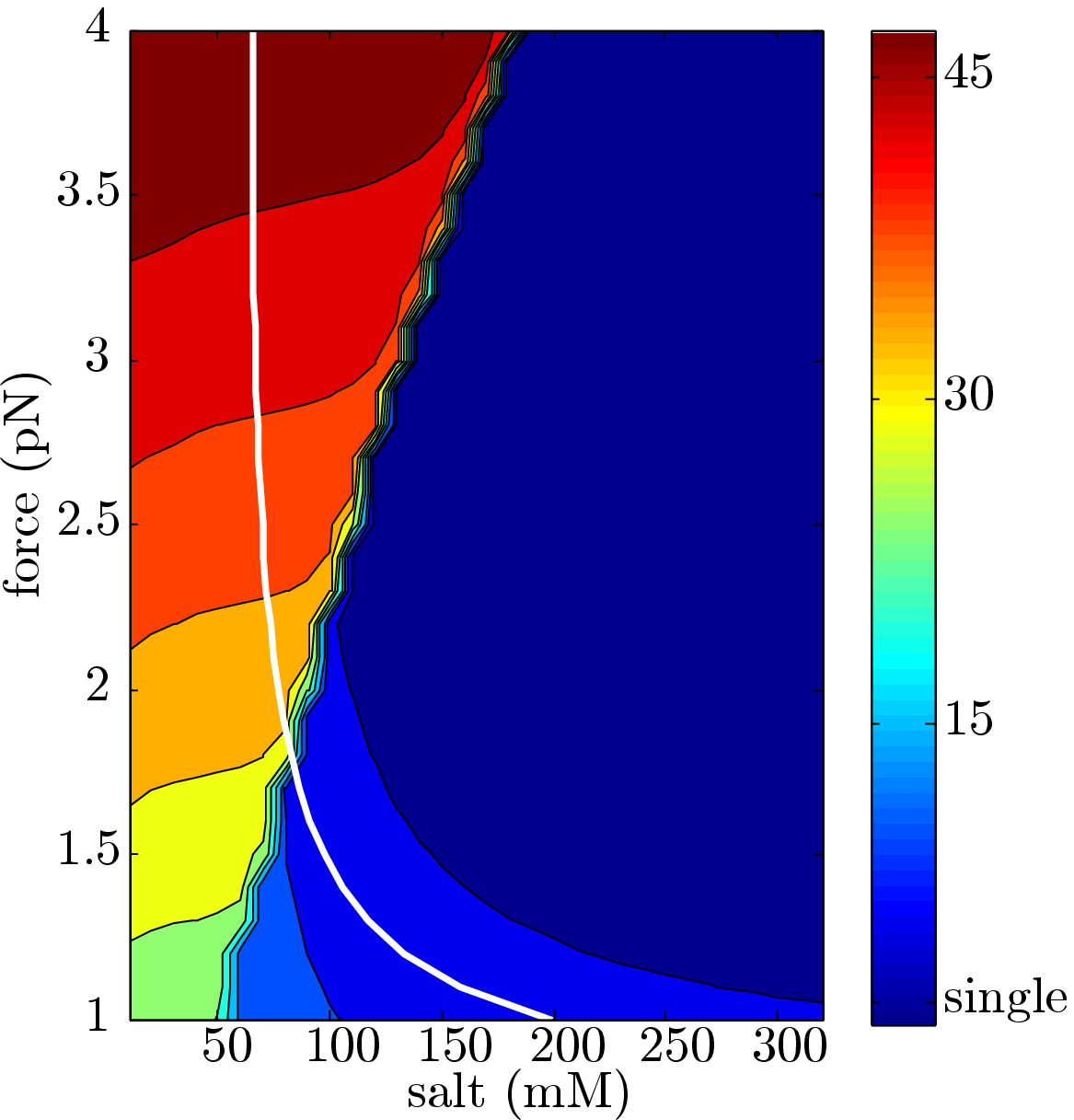}\label{fig:m1400proj}}
\caption{Maximal number of plectonemes as a function of tension and salt concentration for a chain length of $1400$ nm. The $3$-d plot~(a) shows how the number of plectonemes goes down with decreasing force at low salt but increases at high salt. The anomalous behavior at low salt is a consequence of the growing loop size limiting the maximal number of loops that fit on the chain. The wiggles are an artifact of the interpolation used. The contour plot of the same data~(b) shows a clean border  between low and high number. The white line is the $\zeta=1$ -line, that marks the border of single and multi-plectoneme behavior. The difference between the two is caused by the growing number of plectonemes at the end of the plectoneme slope, while $\zeta$ is a measure for the main part of the slope}
\label{fig:finitechain}
\end{figure}

We have seen how the appearance of multiple plectonemes can influence the turn extension curve, but it is often masked by a value of $r_{\wri}$ close to one. Luckily there is another handle through the torque to which we will come back in the next section. But even the torque behavior after the transition does not necessarily change with the onset of multiple plectonemes. There is yet another property that does always change at the moment that the plectoneme density increases. This is caused by the aforementioned fast twist diffusion: two plectonemes can exchange length through twist mediated diffusion which is expected to be much faster than any single plectoneme can diffuse. It also allows for plectoneme diffusion in a crowded environment. This last aspect could be important in vivo where for example a change of tension could regulate the ``capture'' or release of a plectoneme in a pocket within a crowded environment. With this in mind it is interesting to examine the change in the number of plectonemes for a finite chain (Fig.~\ref{fig:finitechain}.
The transition from a single plectoneme to a multi-plectoneme state happens over a narrow band in the tension salt configuration plane. It makes again sense to speak of two separate phases, the normal single plectoneme phase characterized by slowly diffusing if not immobile plectonemes and a multi-plectoneme phase where plectonemes can diffuse even in crowded environments.  It has to be kept in mind that this maximal number of plectonemes might occur only at the end of the plectoneme slope. This is especially true for conditions where $r_{\wri}>1$. For this reason $\zeta$ might give a better handle on the mobility of plectonemes.
It is interesting to note that  $\Delta=0$ plays a key role in the properties of the supercoiling chain like it did in the zero temperature chain, cf. Eq.~\eqref{eq:pleclength}. The effect of the writhe ratio $r_{\wri}$, a minor one  on $\zeta$ but a major one on the maximal loop density, is new and of entropic origin.

\section{\label{sec:comp-with-exper}Comparison with experiments}

To test the validity of the model over an extensive range of parameters, use has
been made of a series
of measurements performed by the Seidel group in Dresden. For combinations of
forces from \SIrange{0.25}{4}{\pico\newton} and salt
concentrations from \SIrange{20}{320}{\milli\Molar} the turns extension curves
were measured for chains of approximately \nm{600} contour-length. We smoothed the
experimental data with a moving average algorithm. To
correct for the geometry of connection to the beads and substrate, the effective
contour length of the chain has been
obtained by fitting the $0$ turns extension to the ideal not torsional
restricted worm like chain. Up to lowest order
this should be equivalent to the torsionally constrained $0$ turns
configuration. The effective chains thus obtained
have a length that varies between \nm{570} and \nm{630}. A set of measurements
under varying forces, but constant salt
concentration has been performed on one chain allowing us to verify that the
effective chain length stays more
or less constant once the geometry of the chain attachment is fixed. Only for
forces below \SI{1}{\pico\newton} the effective chain length decreases. This is
partly due to the bent chain attachment, combined with too wildly fluctuating
chains for our perturbative model.

The minimization procedure was initiated as follows: starting from the
bifurcation point, $\lk_{cr}$, the applied linking number per length was set to
$\nu=\lk_{cr}+0.2$ to assure the linking number density is far after the transition. The parameters of the model were set to $\lk=0.8 \lk_{cr}$,
$\alpha=1$, $\sigma_r=1/(2\kappa)$, and $R=1+\kappa^{-1}$ (in nm such that the potential of a cylinder with a radius of $1$ nm is in the linear regime at $R$). The free energy for a
single chain was minimized after which the obtained values were used to set
$\nu$ to $(\wri+\lks+2\lk_{cr})/2$, setting the linking number density halfway between the bifurcation point and the maximum. The resulting plectoneme parameters were used as starting values for another minimization. In that way the applied linking number
is approximately halfway in between the critical value and the maximal value.
The reasoning is that with a linking number close to the bifurcation point the
influence of an incomplete description of the end loop becomes too strong, while
a linking number too far from the transition might underestimate the influence
of multi plectoneme configurations. The precise value is not very important. Too
close to the bifurcation point the chain collapses before the transition in
low salt condition. The reason is not so much the influence of the loop but a
$K^2$ value that gets too
low. Of course any prediction based on the model for $K^2$ values below $3$ is
unreliable.
The generation of the force extension behavior is based on plectoneme energies
from this minimization. The whole procedure is very fast.

\begin{figure}
  \includegraphics[width=0.75\textwidth]{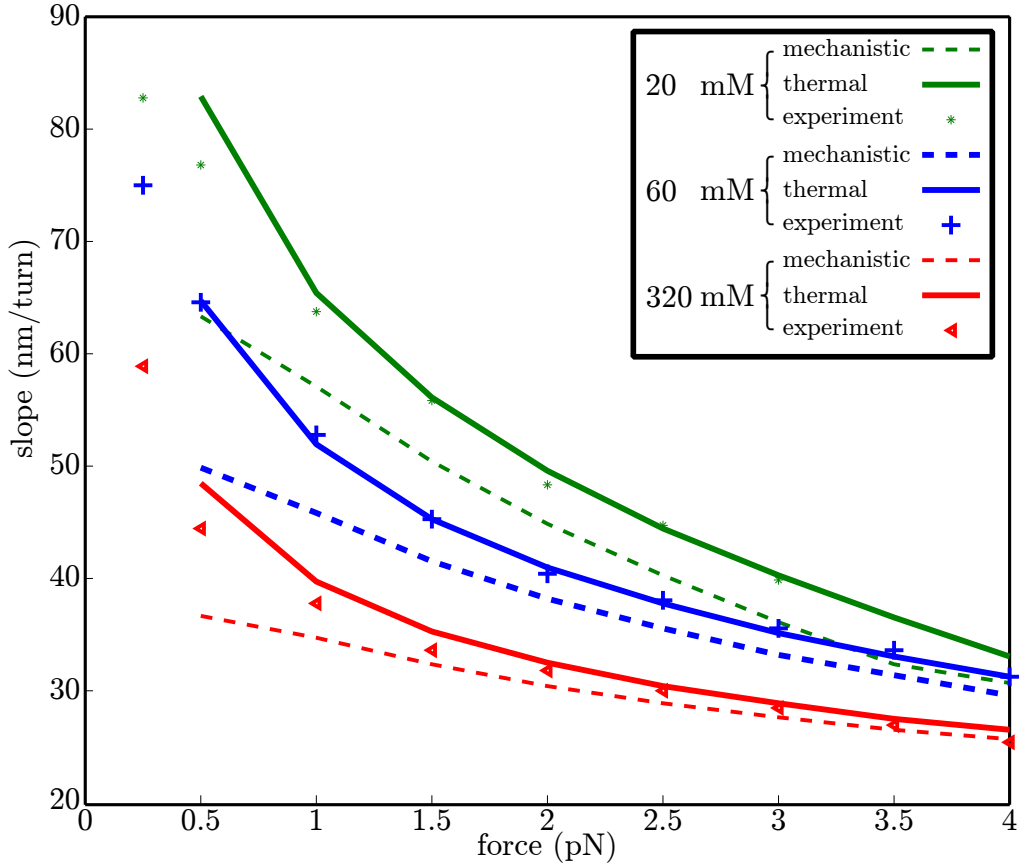}
\caption{Slopes with and without
thermal contributions. The dotted curves were calculated from the model up to
section~\ref{sec:thermal}, equation~\eqref{eq:freeplect}.
The solid thermal curves include multi-plectonemes and were calculated using the
method
outlined in the text.}\label{fig:totslope}
\end{figure}

\begin{figure}
\centering
\subfloat[]{
  \includegraphics[width=0.47\linewidth]{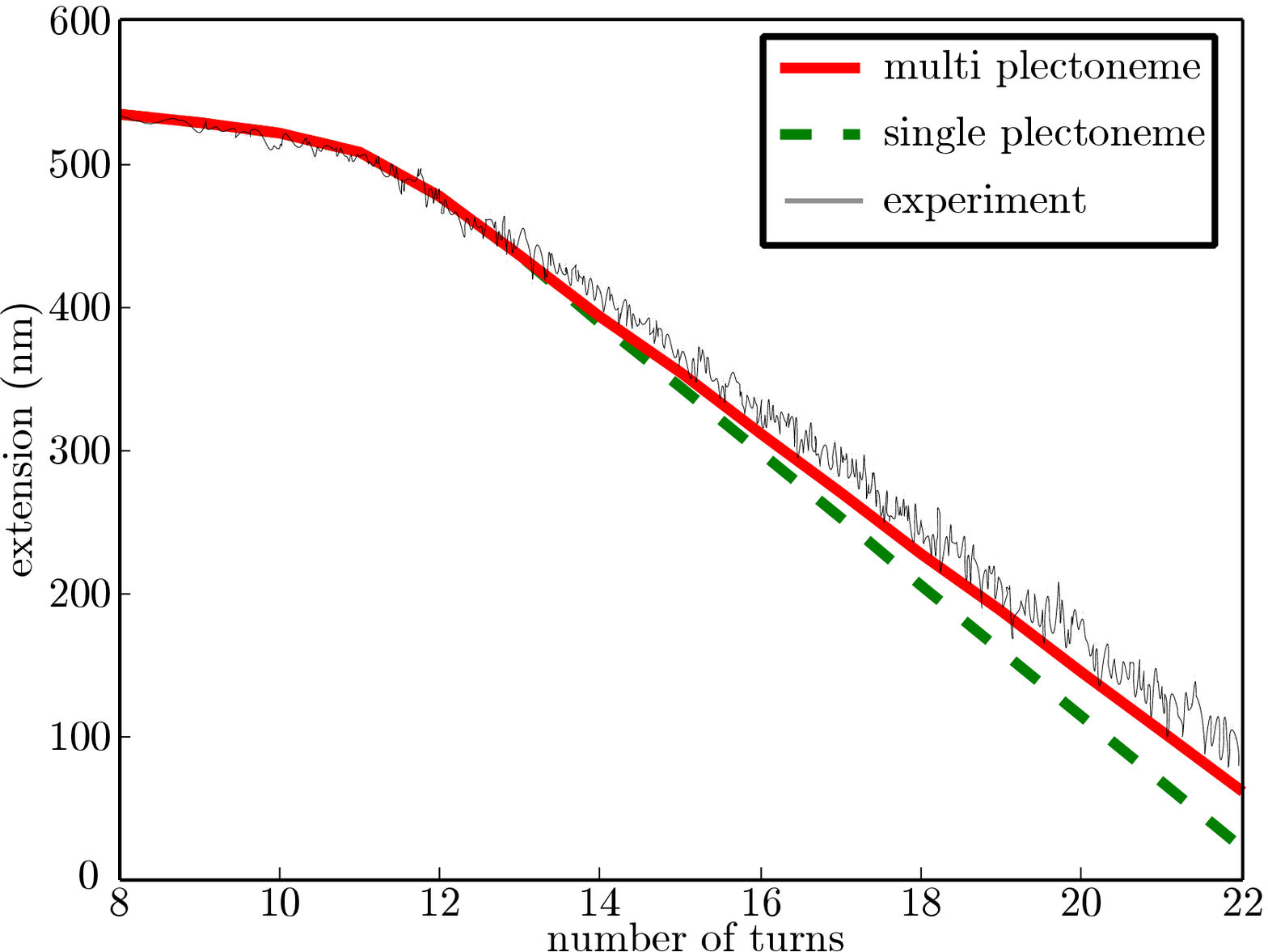}
\label{fig:singmult}}
\quad
\subfloat[]{
\includegraphics[width=0.47\linewidth]{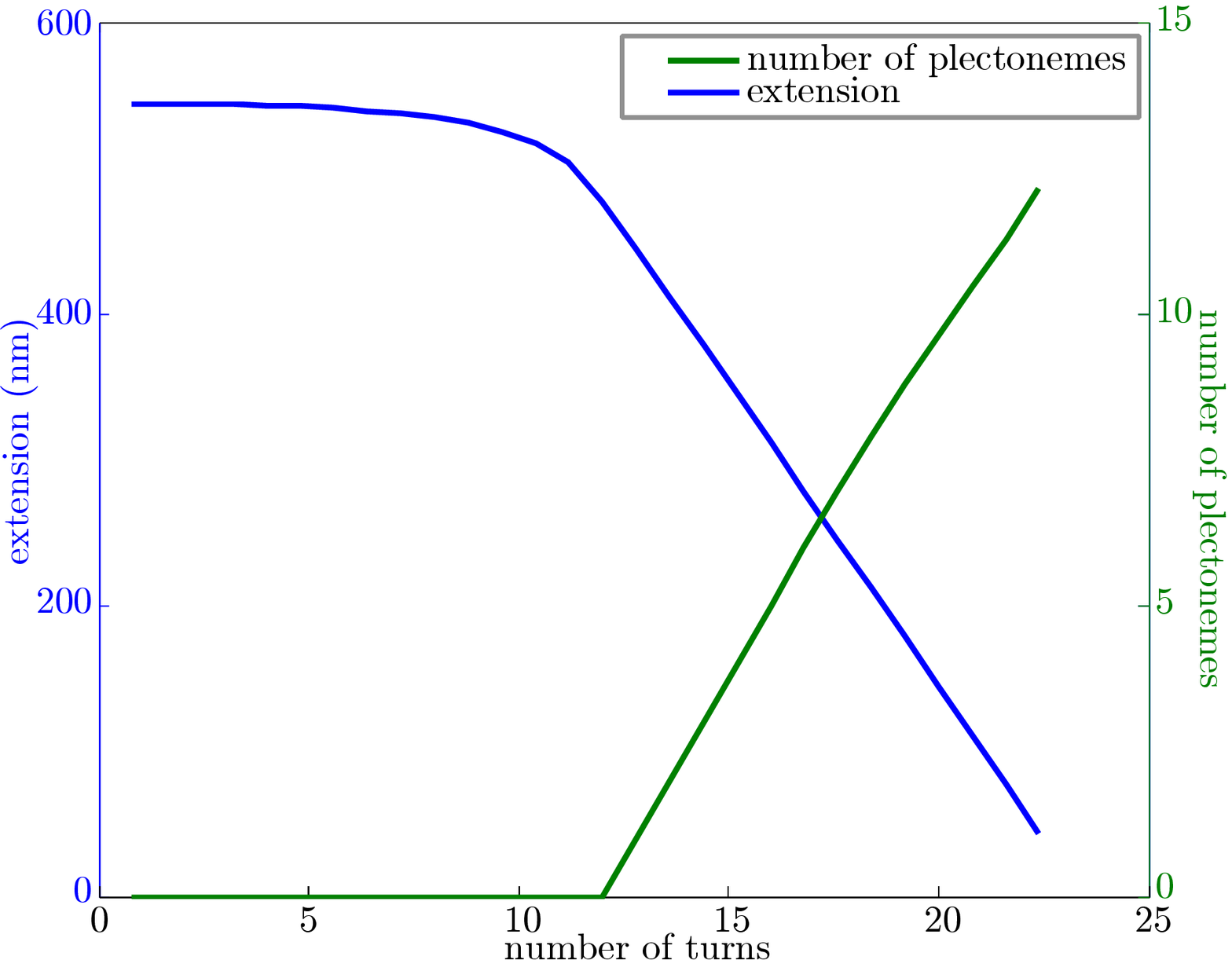}
\label{fig:plectnum}}
\caption{Influence of multi-plectonemes on the turns extension plot at low salt. For a chain with a contour length of \SI{600}{\nm}, a tension of   \SI{3}{\pico\newton} at a \SI{20}{\milli\Molar} salt concentration there is a noticeable effect on the slope~(a). The growing number of plectonemes and not a growing  plectoneme length is responsible for the slope~(b). The experimental data are from the Seidel lab.}
\label{fig:mpexample}
\end{figure}

As a first test of our model we compare predicted plectoneme slopes to those
determined in experiments. Note that the choice of where to measure the slope is
not always obvious in both theory and experiment. Whenever
there was a clear constant slope visible it was taken as the slope, otherwise
the first slope after the transition was taken. Especially for the short
\nm{600}
chains it was not always clear what to take as slope. This is especially true
for low salt, \SIrange{20}{60}{\milli\Molar}, conditions. Nonetheless the slopes
for the full range indicated a nice agreement between experiment and model. The
results for $20$, $60$, and $320$ \si{\milli\Molar} are in
Fig.~\ref{fig:totslope}.
The influence of the multi plectoneme phase is clearly visible for low salt
concentrations. There is also a clear improvement in the low force range,
although
there the value of $K^2$ of $2$ or lower around the transition point makes the agreement
mere coincidental.
The turn-extension plot at \SI{20}{\milli\Molar} and \SI{3}{\pico\newton} in
Fig.~\ref{fig:singmult} shows the details. The transition happens
at
a lower linking number than in the experiment, presumably because it is too
close to the bifurcation point for a reliable perturbative calculation. To produce the plots the torsional persistence
length was lowered to \nm{90} from \nm{110} to get the transition
point close to the experimental value.
In Fig.~\ref{fig:plectnum} the number of plectonemes is set out against the
number of turns for these conditions.

\begin{figure}[htp]
\centering
\subfloat[] {
  \includegraphics[width=0.46\linewidth]{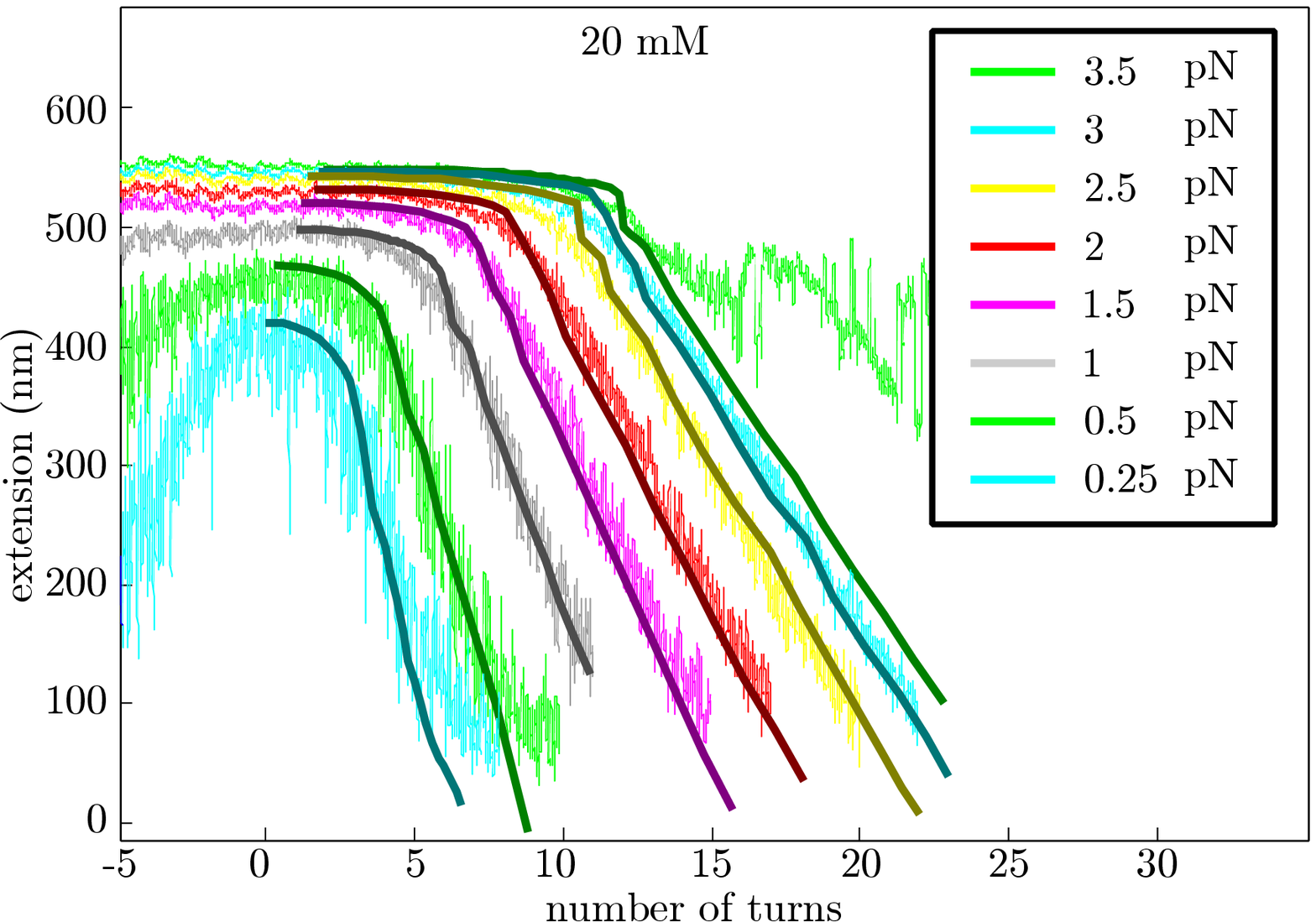}\label{fig:20mM}}
\quad
\subfloat[]{
   \includegraphics[width=0.46\linewidth]{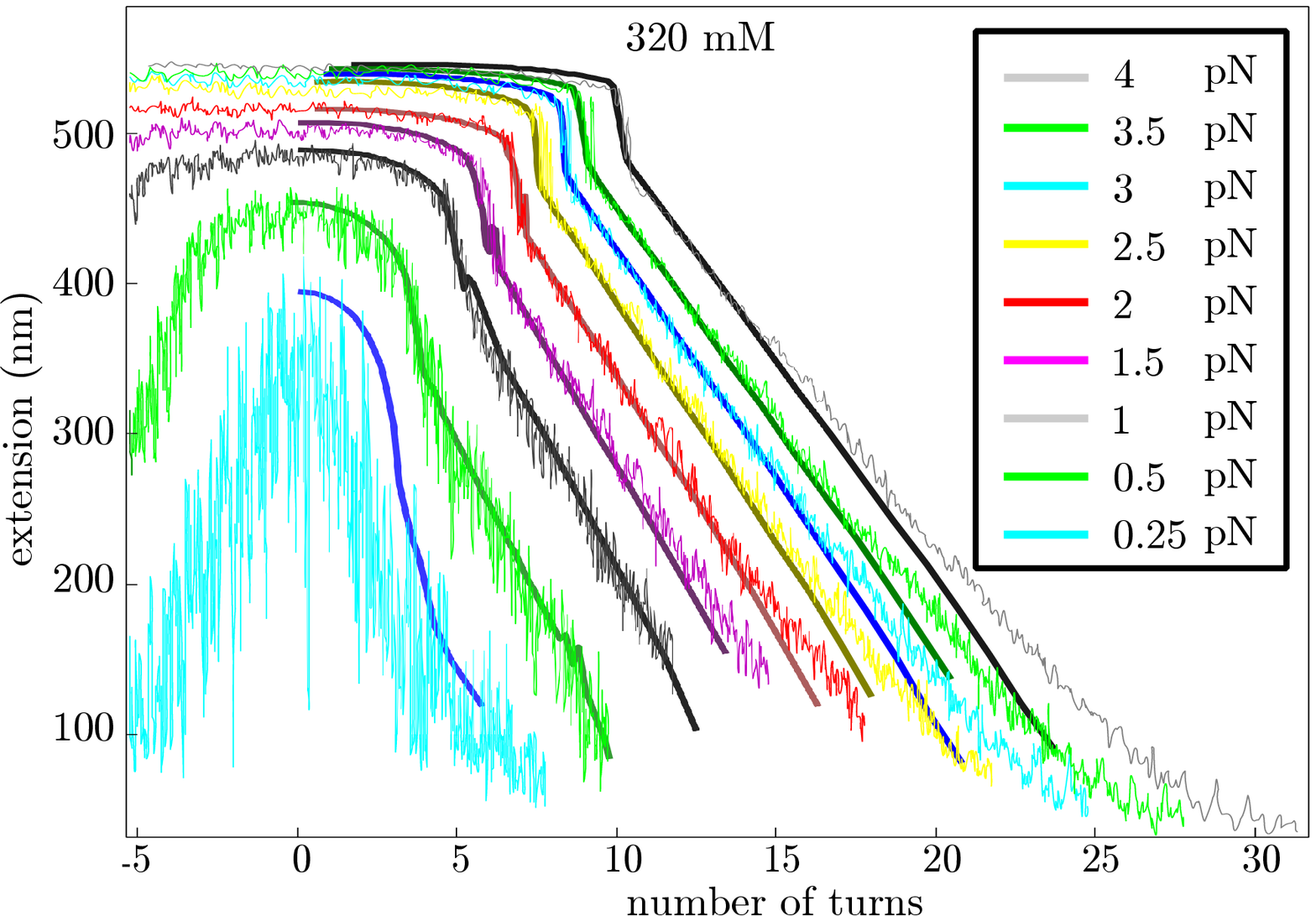} \label{fig:320mM}}
\caption{Turns versus
extension plots for \SI{20}{\milli\Molar}~(a) and \SI{320}{\milli\Molar}~(b) monovalent salt concentrations under varying tension. The torsional
persistence length for most salt gave the best fit  for \nm{110}.  For  \SI{20}{\milli\Molar} a lower value of \nm{90} had to taken to get an almost perfect agreement, but as explained in the text it
might be a calculational artifact due to the proximity of the
bifurcation point.}
\label{fig:highlow}
\end{figure}

The curves for \SI{20}{\milli\Molar} and \SI{320}{\milli\Molar} are shown in
Fig.~\ref{fig:highlow}. For most cases our model
predicts the experimental curves well.
The behavior at the transition at \SI{20}{\milli\Molar} and a tension above \SI{3}{\pico\newton} is not well defined perturbatively, since the straight solution has a $K^2$ value below $3$ before the first plectoneme solution becomes available.
The \SI{20}{\milli\Molar} measurements show an exceptional behavior at
\SI{3.5}{\pico\newton}. It is possible that the chain undergoes a phase
transition as has
been suggested\cite{Maffeo:2010}. Another possibility is that because
plectoneme formation is relatively expensive, starting plectonemes are extremely
unstable. That can explain the sawtooth behavior with signs of attempts at
plectoneme nucleation.

\begin{figure}[htp]
\centering
\subfloat[Turns versus extension
plots for a chain of \nm{3540} with tension from \SIrange{1}{4}{\pico\newton} in
\SI{320}{\milli\Molar} salt. A torsional persistence length of \nm{120}
was used.]
  {\includegraphics[width=0.45\textwidth]{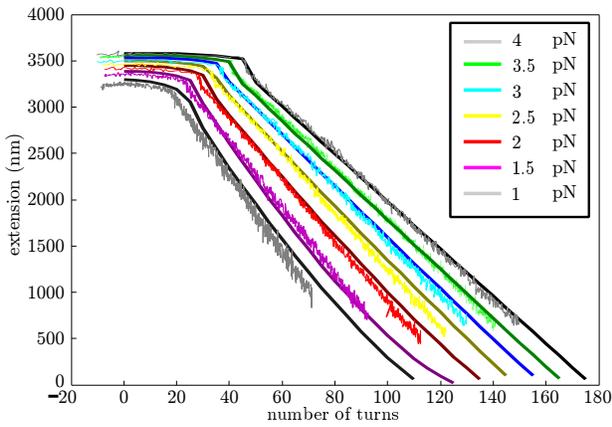}
  \label{fig:long}}
  \quad
  \subfloat[Some of the magnetic tweezer measurements
from Mosconi et al.\cite{mosconi2008torque} with a \SI{5.4}{\micro\meter} long chain in $100$ mM salt for $3$ different forces.  The curves are from our model with a torsional persistence length of $115$ nm]
 { \includegraphics[width=0.45\linewidth]{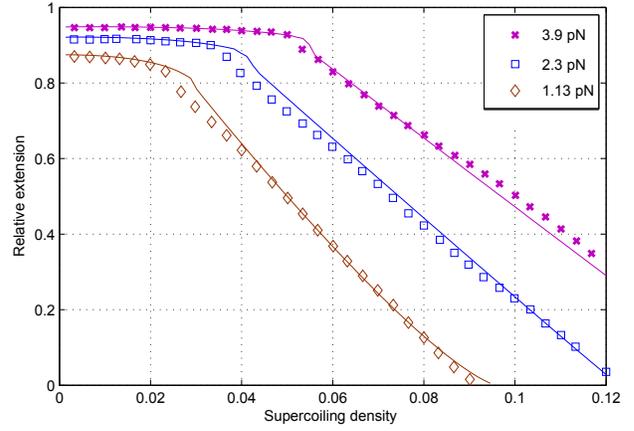}\label{fig:mosccomp}}
  \caption{Long chains}\label{fig:longchains}
\end{figure}

A set of experiments performed on a \nm{3850} chain
in a \SI{320}{\milli\Molar} solution with the same setup shows a longer clear
slope in Fig.~\ref{fig:long}. The transition point suggests here a \nm{120}
torsional
persistence length.

Another test of the model is the analysis of the plectoneme torques. The torque
is obtained by dividing the increase of the free energy by the
rotation angle that caused it.
It is commonly believed that the linear slope of the curves coincides with a
state of constant torque\cite{strick2000twisting,Marko:2007}. This makes it
attractive to use the DNA plectoneme as a source of constant torque in the study
of molecules that interact with DNA like topoisomerase and helicase. One way to
measure this plectoneme torque is by using a specially nano-fabricated quartz
cylinder in conjunction with an optical tweezer\cite{deufel2007nanofabricated}.
The setup seems to be very promising enabling the measurement of torque at the
same time as force and extension. A small set of measurements were done with
relatively short chains of \nm{700}\cite{Forth:2008}.
Another method makes use of the
constant torque in the plectoneme region combined with Maxwell relations
between torque/linking number and force/extension as free energy parameters.
The method calculates the plectoneme torque over a large range of forces using
an approximately linear linking number/torque relation before the transition at
high tensions. Assuming a
constant torque after plectoneme formation, the torque for a large range of data
can be calculated just from the turn extension plot. This is the setup from
Mosconi et al.\cite{Mosconi:2009}. The resulting torques in the two types of
measurements\cite{Forth:2008,Mosconi:2009} seem
to differ. It could be that the salt concentrations differ too
much, or that the response of the optical trap is too slow.
\begin{table}[t]
\centering
 \caption[Indirect torque measurements using Maxwell relations,]{Indirect torque
measurements using Maxwell relations\cite{Mosconi:2009} compared to the
theoretical values from our model}\label{tab:torque}
\begin{ruledtabular}
\begin{tabular}{c c c c }
 Salt (\si{\milli\Molar})& Force (\si{\pico\newton}) & Exp. Torque(\si{\pico\newton\nm})\cite{Mosconi:2009} & Theoretical Torque (\si{\pico\newton\nm})\\ [0.5ex]
\hline\\[-1.5ex]
 10 & 2.86 & 28.1 &35.0 		\\
     & 2.53 & 26.2 & 32.0 	\\
50 & 3.66 & 29.6 & 34.7 	\\
    &  3.23 & 27.4 & 32.4 	\\
100 &   3.33& 24.4 & 30.1	\\
     & 2.61 & 20.7 & 26.3   	\\
500 & 4.33 & 22.3 & 29.6 	\\
      & 3.80 & 20.2 & 27.5
\end{tabular}
\end{ruledtabular}
\end{table}
It is interesting to compare the torques that our model predicts with those of Ref.~\cite{Mosconi:2009}. To our surprise the torques we calculate differ from
their measurements substantially enough to doubt the validity of our model, see
table~\ref{tab:torque}. The torque from the model was calculated just after the transition at the start of the plateau by calculating the change in free energy as function of the change in linking number. This deviation in torque came not totally unexpected, since the
torque data were one of the reasons for Maffeo et al\cite{Maffeo:2010} to
incorporate a charge reduction factor into their model. What is somewhat
mysterious is that the force extension curves themselves are in good agreement with our model as illustrated by Fig.~\ref{fig:mosccomp}.
If the torque only depends on the shape of that curve,  while Maxwells relations hold per definition, somewhere a wrong
assumption must have been made.

\begin{figure}[htp]
\centering
\subfloat[Torque as function of supercoiling
density for three different tensions calculated from the model compared to the
indirect determination of the torque using Maxwell relations under constant
plectoneme torque assumption\cite{mosconi2008torque}. The conditions are the
same
as in Fig.~\ref{fig:mosconi}]{
    \includegraphics[width=0.44\linewidth]{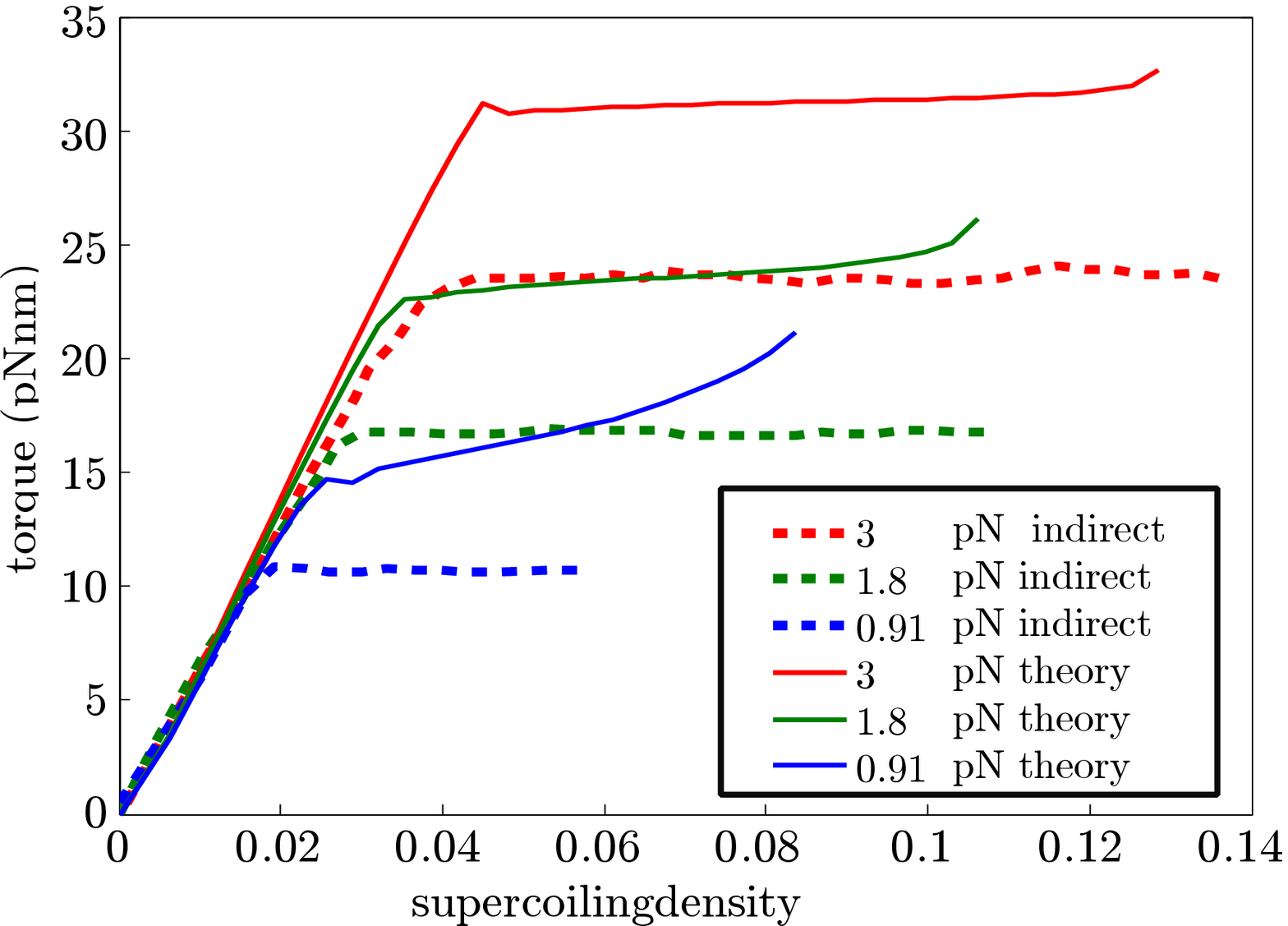}
 \label{fig:mosctorque}}
\quad
\subfloat[Comparison between
experimental results from an optical tweezer
experiment using a quartz cylinder to directly measure the
torque\cite{Forth:2008} and our theoretical model. The DNA molecule has a
contour length of \nm{725}. The monovalent salt concentration is
\SI{150}{\milli\Molar}]{
   \includegraphics[width=0.44\linewidth]{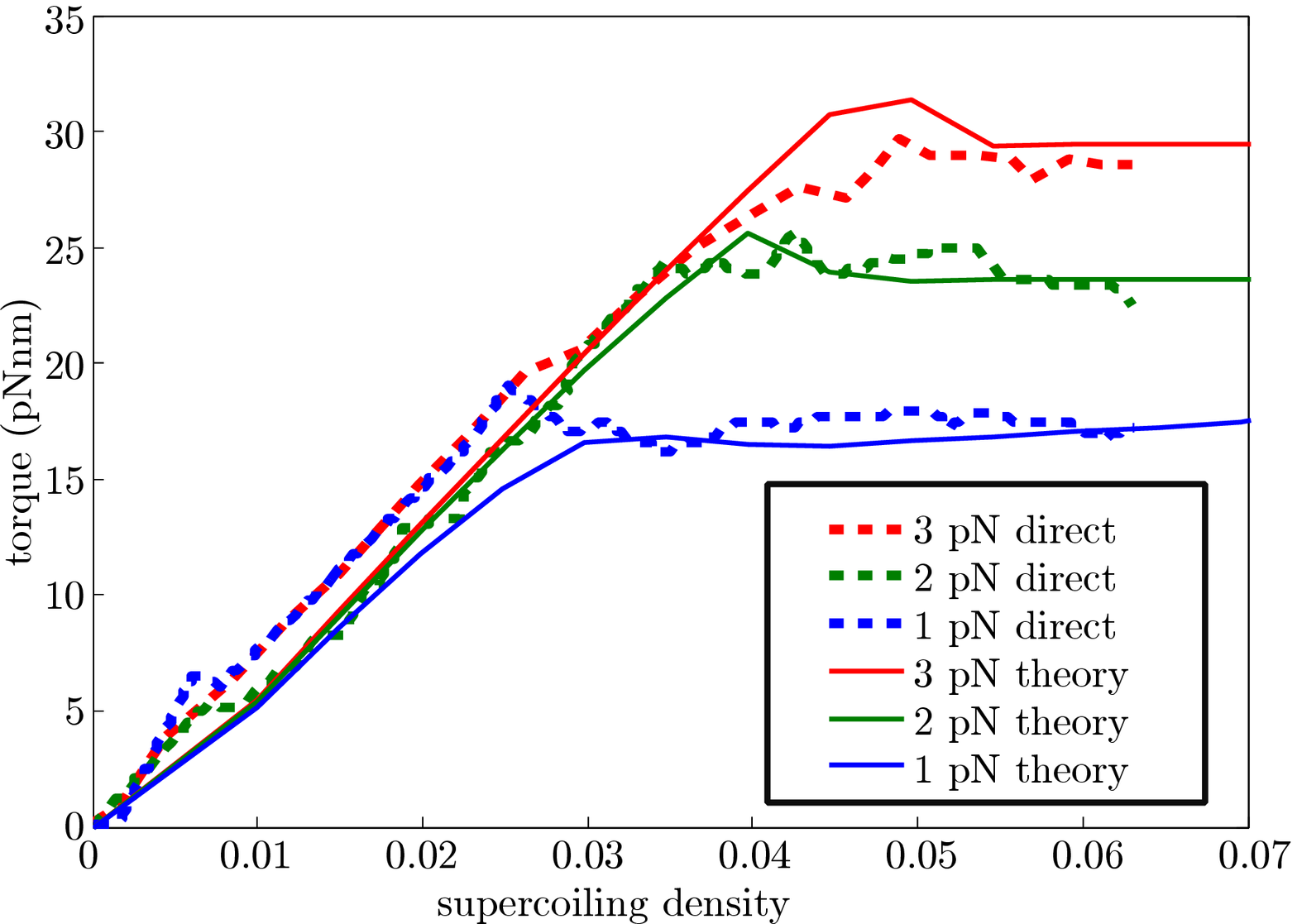}
 \label{fig:forthtorque}}
\caption{}\label{fig:thetorques}
\end{figure}

\begin{figure}[htp]
    \includegraphics[width=\linewidth]{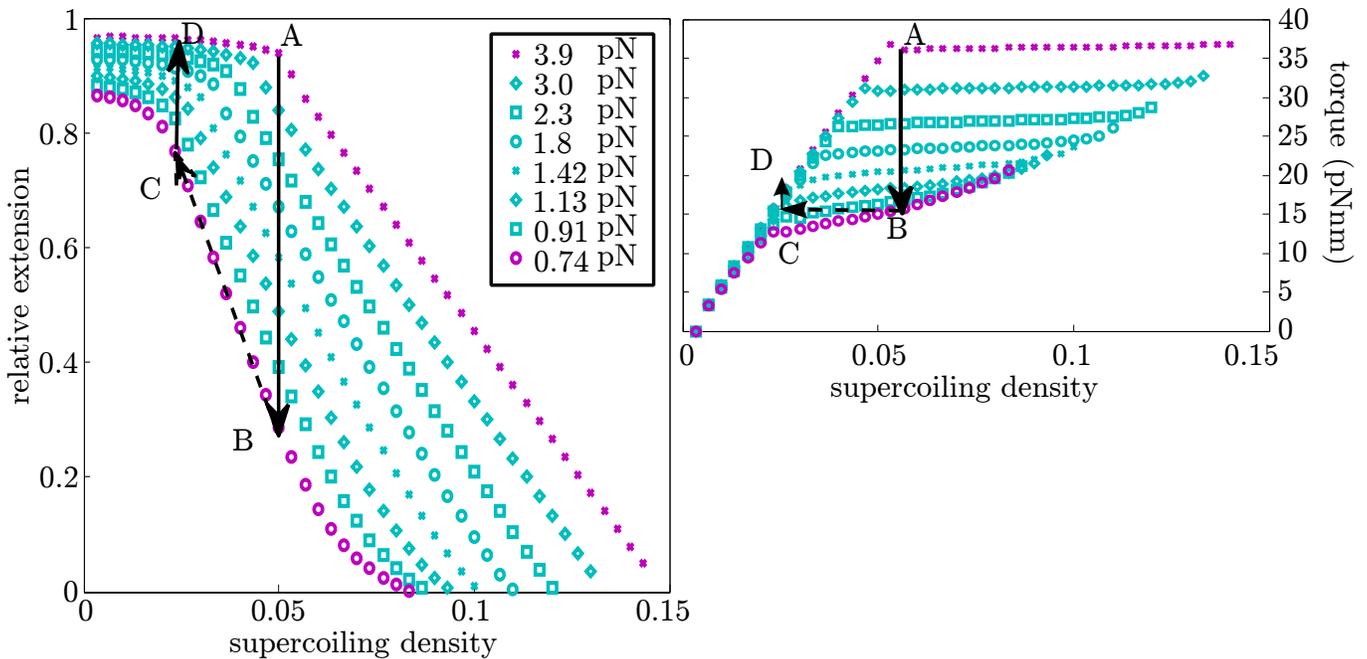}
  \caption{Illustration of the magnetic tweezer measurements
from Mosconi et al.\cite{mosconi2008torque} as basis for indirect torque
measurements. The curves are
calculated from our model for a range of tensions where fluctuations are small enough
using the criterion $K^2\geq 3$. The contour-length is \SI{5.6}{\micro\meter},
monovalent salt concentration of \SI{100}{\milli\Molar}. On the left are the resulting turn extension plots. On the right are torques from our model. Notice how the torque is increasing in the plectoneme region caused by the growing number of plectonemes. The circuit ABCD results in too high a torque at C when one assume the plectoneme torque to be constant. }
\label{fig:mosconi}
\end{figure}
Comparing our torque predictions with the direct torque measurements from the
older optical tweezer measure\-ments\cite{Forth:2008} reveal however a remarkable
good agreement as is shown in Fig.~\ref{fig:forthtorque}, where the torques are
shown as a function of the supercoiling density defined as the ratio of the
linking number density to the linking number density of the two strands of the
double helix when the chain is straight and relaxed. This last density is of
course $1/$helical repeat $=1/3.6 \text{ nm}^{-1}$.

The culprit is readily revealed as the multi-plectoneme phase. In extracting the
torque from the force extension measurements an essential assumption is that the
torque in the linear slope is constant. That almost presupposes that
the slope is a one plectoneme slope. Lacking a method to verify this assumption
it had to be accepted on face value. In reality the torque is not constant at
all for lower forces. Thanks to the fast increasing number of plectonemes along
the chain the torque is almost linearly \emph{increasing} invalidating the
calculations. When we take this increase into account, the resulting torque
values agree
again wonderfully well with the predictions from our model.

As an example we borrow the calculations from Mosconi et
al.\cite{mosconi2008torque}. The relevant curves are in
Fig.~\ref{fig:mosconi}. The Maxwell relation
calculations are performed over the path as shown in the figure on the left. The
resulting torque for \SI{3.67}{\pico\newton} is \SI{27}{\pico\newton\nm}. But
if we examine
the torque as calculated from the model the result is higher, around
\SI{34.9}{\pico\newton\nm}. Though the torque is constant for the high-tension
slope,
the path from B to C in Fig.~\ref{fig:mosconi} is one of decreasing torque
thereby resulting in a
too low estimate for the plectoneme torque.

Finally inspired by our preliminary results an experiment was setup, where the first results\cite{loen:2012} became recently available. Multi-plectonemes were visualized under conditions were we had expected them to exist. Also the fast twist mediated diffusion, only possible when at least two plectonemes are present, was observed. The resolution is at this moment not good enough to extract plectoneme number densities.

\section{\label{sec:close}Comparison with other models and outlook}
Over the years numerous models have been proposed, all of them bringing in some new ideas. We cannot compare our model against all of them, but we will discuss some recent works that are of interest with respect to our model.

First of all there is the model proposed by the Seidel group\cite{Maffeo:2010} where, based on a non-thermal model like the one in section~\ref{sec:1}, it was proposed that the charge of the DNA molecule should be reduced by a factor with a value determined by the experimental slopes. The reasoning was that part of the counter ions might be confined to the grooves of the double helix. An assumption was that the thermal shortening of the DNA would be the same in the plectoneme and in the tails. Finally the resulting potential was used as basis of Monte Carlo simulations that confirmed that the slopes were unaffected by fluctuations.
Now there exist a couple of objections. The lack of any influence of thermal fluctuations is hard to understand, the way fluctuations are restricted being quite different in tails and plectoneme. It might be that the $5$ nm segment length chosen for the MC simulations is too large for capturing the essential part of the spectrum.
Perhaps more problematic is the charge reduction of more than a half.  It would mean that the concept that DNA is a strong polyelectrolyte with respect to the phosphates, one charge per base pair, is wrong. This would contradict direct experimental evidence (for example Ref.~\cite{leikin:1991}), but also indirect measurements e.g. concerning the pressure of viral DNA, DNA condensation models and more. It is puzzling how the inhomogeneity  of small monovalent counter ions far within the inner layer can affect the potential outside of the nonlinear domain. Furthermore as we have shown the slopes are in fact not that dramatically affected since the writhe ratio is close to one.
Another more recent work introduced small loops in addition to the possibility of multiple plectoneme formation\cite{marko:2012}. The resulting modeling can not faithfully reproduce the measured curves though. One problem is that these little loops do not have any electrostatic or entropic repulsion incorporated. It is possible to add a more detailed description  to these loops, as one of us has shown\cite{Emanuel:2010}, but then it is only a small step to acknowledge that these loops are in fact zero length plectonemes.

In this paper we have for the first time developed a  model  for plectoneme formation that makes a consistent description of thermal fluctuations on all scales. The model is perturbative which has its limitations, but under  conditions that a perturbative treatment makes sense it performs well.
We discovered a sharp boundary between single plectoneme and multi-plectoneme behavior and argued for possible biological implications. This multi-plectoneme behavior at the same time resolves a couple of anomalies in experimental data.
There are of course still many open questions. We think that the most pressing concerns the direct measurement of torques over the full plectoneme slope. This could shed some light on the correct cutoff value. A full analysis of the length drop at the transition we will leave for a future publication.

\begin{acknowledgments}
This research would have been impossible without the excellent sets of measurements provided by the group of Ralf Seidel. We are thankful for the inspirational discussions with Theo Odijk and Andrey Cherstvy. This work would not have been possible without the support from Gerhard Gompper and the Forschungszentrum J\"ulich.
We acknowledge the many discussions with Marijn van Loenshout and Cees Dekker and are glad that our theoretical predictions helped them to find the multi-plectoneme phase experimentally.
\end{acknowledgments}
\appendix
\section{\label{app:writhe}Writhe of a plectoneme}
In principle the writhe of the plectoneme can be calculated using Fuller's
equation and
continuity. Care should be taken since the plectoneme moves through a curve
with an anti-aligned tangent once every full turn of the plectoneme, when one
considers the (un)winding as the homotopy to the straight line.
Since we intend to use an exact expression for the writhe at least for the
ground state it is instructive first to
calculate the writhe density for the plectoneme using Fuller's equation with
respect to the plectoneme-axis for both strands, forgetting loop and tail:
\begin{align}
  \wri_1(s)&=\frac{1}{2\pi}\frac{\cos(\alpha)(\sin(\alpha)-1)}{R(t)} & s&\in
[0,l_p/2] \nonumber \\
 \wri_2(s)&= \frac{1}{2\pi}\frac{\cos(\alpha)(\sin(\alpha)+1)}{R(t)} & s&\in
[l_p/2+l_l,l_p+l_l].
\end{align}
We could in a hand waving fashion define an ``average'' writhe density as
\begin{align}
 \wri(\alpha,t) &\stackrel{?}{=}
\frac{1}{2}(\wri_1(s)+\wri_2(l_p+l_l-s))=\frac{
\cos(\alpha)\sin(\alpha)}{2\pi R(t)}
\end{align}
The problem is that this definition, giving the usual relation, is based on
Fuller's equation with respect to another axis than we started with, and we have
not taken
the writhe of the end-loop into account.

A correct way that shows the importance of the rotation of
the closing loop is to use also here the $z$-axis as reference. Opposing points
on the plectoneme strands have in this case the same writhe:
\begin{align}\label{eq:writhebare}
\wri_b(s)&=\wri_b(l_p+l_l-s)= \frac{1}{2\pi
}\frac{\sin\alpha\cos\alpha}{R(t)}\left[1-\frac{1}{
1+\cos\alpha\cos\left((s_0+s)\frac{\cos\alpha}{R(t)}\right)}\right]
\end{align}
A surprising $s$ dependence enters the writhe density of the plectoneme. The
subscript $b$ is as a reminder that this is a bare writhe density that does not
include interactions with the rest of the chain.
Adding plectoneme length also changes the writhe of the end-loop though.
The closing end-loop is described at the onset of the plectoneme formation by
some space
curve $\vec{r}_0(u)=(r_x(u),r_y(u),r_z(u)), u\in[0,l_l]$, with boundary
conditions: $\vec{r}_0(0)=\vec{r}_p(0)$ and $ \vec{r}_0(l_l)=\vec{r}_p(l_l)$. We
furthermore assume the connection between the plectoneme and the end-loop to be
smooth, making the tangent well defined at the boundaries. The increase of
the plectoneme by an amount of contour length $2s$ causes the end-loop
to rotate around the $x$-axis by an angle $\phi(s)=s\cos\alpha/R(t)$. The rotated loop is given by:
\begin{align}
 \vec{r}_s(u)=\hat{O}_x(\phi(s))\vec{r}_0(u)=\left(\begin{array}{c}
                                                                 r_x(u)\\
\cos\phi(s)r_y(u)+\sin\phi(s)r_z(u)\\
-\sin\phi(s)r_y(u)+\cos\phi(s)r_z(u))
     \end{array}\right)
\end{align}
This rotation induces an $s$ dependent change in the writhe of the loop to:
\begin{align}
\begin{split}
 \Wrll(s)&=\\
\frac{1}{2\pi}&\int_0^{l_l}\de u
\left(\frac{\cos\phi(s)(t_x(u)\dot{t}_y(u)-\dot{t}
_x(u)t_y(u))}{1-\sin\phi(s)t_y(u)+\cos\phi(s)t_z(u)} -\frac{\sin\phi(s)(t_z(u)\dot{t}_x(u)-\dot{t}_z(u)t_x(u))
}{1-\sin\phi(s)t_y(u)+\cos\phi(s)t_z(u)} \right).
\end{split}
\end{align}
The superscript is just a reminder that it is not the writhe of the full
homoclinic solution, but just of that part that detaches to function as end loop
for
the plectoneme.
Note that this writhe is not necessarily well defined. In fact since the length
of the loop is finite, its $x$-component is bounded and thus has at least one
point where the tangent lies in a plane perpendicular to the $x$-axis. This
tangent will be once every full turn of the plectoneme antipodal to the
z-axis and thus invalidates Fuller's equation.

We can nonetheless calculate the differential change of this writhe per
plectoneme contour:
\begin{align}
\begin{split}
 \frac{\de \Wrll}{\de s}=
\frac{\cos\alpha}{2\pi R(t)}&\int_0^{l_l}\de
u\frac{-\dot{t}_x(u)-\sin\left(s\frac{
\cos\alpha}{R(t)}\right)(t_x(u)\dot{t}_y(u)-\dot{t}_x(u)t_y(u))} {
\left(1-\sin\left(s\frac{
\cos\alpha}{R(t)}\right)t_y(u)+\cos\left(s\frac{\cos\alpha}{R(t)}
\right)t_z(u)\right)^2}\\
&\qquad-\frac{\cos\left(s\frac{
\cos\alpha}{R(t)}\right)(t_z(u)\dot{t}_x(u)-\dot{t}_z(u)t_x(u)) } {
\left(1-\sin\left(s\frac{
\cos\alpha}{R(t)}\right)t_y(u)+\cos\left(s\frac{\cos\alpha}{R(t)}
\right)t_z(u)\right)^2} \\
= \frac{\cos\alpha}{\pi R(t)}&\frac{t_x(0)}{1-\sin\left(s\frac{
\cos\alpha}{R(t)}\right)t_y(0)+\cos\left(s\frac{\cos\alpha}{R(t)}\right)t_z(0)}
,
\end{split}
\end{align}
where use has been made of the unimodularity of the tangent vector and its
symmetry: $t_x(0)=-t_x(l_l), t_{y,z}(0)=t_{y,z}(l_l)$. Making use of the
boundary conditions we finally find
\begin{align}
 \frac{\de \Wrll}{\de s}&=\frac{\cos\alpha\sin\alpha}{\pi
R(t)}\frac{1}{1+\cos\alpha\cos\left((s_0+s)\frac{\cos\alpha}{R(t)}\right)}
\end{align}
By adding this differential writhe density to the ``bare'' writhe density of the
plectoneme as given by Eq.~\eqref{eq:writhebare} (half of it to each strand) we recover
the standard writhe density of a plectoneme~\eqref{eq:plectwrithe}, but now with the
added bonus that the remaining writhe of the end-loop is independent of the
length of the plectoneme. Since it is only in the
end-loop that antipodal points appear along the homotopy, defined by the
explicit formation of the plectoneme, we can state that in this sense the writhe
is additive:
\begin{align}
 \Wr(t,\alpha)&=\Wrl(t) +\Lp\wri(t,\alpha),
\end{align}
with $\Wrl$ and $\wri$ given by Eqs.~\eqref{eq:wrl} and~\eqref{eq:plectwrithe}.

Note that we used implicitly continuity to recover the full writhe of the chain
by adding the differential writhe change of the end loop. In hindsight it is
clear that the end-loop
should be included in the final result. Imagine for example a larger end-loop
such that the helices do not intertwine. The writhe in this case can be calculated immediately without any continuity argument and it is
easy to show from Eq.~\eqref{eq:writhebare} that applying Fuller's equation
to a chain with such a non-intertwining plectoneme of $n$ turns ($l_p=4n\pi R(t)/\cos\alpha$) gives
a writhe of $\Wr-2n$.

\section{\label{ap:fluctuations}Fluctuations of the strands in a plectoneme}
Our treatment of thermal fluctuations in the plectoneme follows largely the work
by Ubbink and Odijk~\cite{Ubbink:1999}, with some catch forced upon us by the
physical conditions. In our case we can not just use
Burkhardts result of the confinement of a rotational relaxed
chain~\cite{Burkhardt:1995}, but have to take the twist along the chain into
account. This has two implications:
$1$. The confinement free energy gets twist dependent corrections, the
calculation of which are presented in Ref.~\cite{Emanuel:2013}.
$2$. The confinement gives a relation between linking number and twist that
depends on the confinement channel width.
This is used to calculate the contour length of the plectoneme in the text.

In this appendix we will discuss how the fluctuations can be separated from the
average plectoneme path.
Thermal undulations effectively shorten the chain within its superhelical path.
This has implications on the bending energy and the writhe density of the
plectoneme. To calculate the effect we attach
to each point along the non undulating path, the $0$-path or $0$-chain, a
triad, consisting of the tangent at that point and two normals. The fluctuations
we can express in deviations in the two normal directions from the $0$-path.
The deviations in the tangential direction follow from the in-extensibility, or
if needed a finite stretch modulus can
be included~\cite{Marko:1998}. For the plectoneme as triad we take its Fresnet
basis,
where the normal is the direction of curvature, which is the radial direction,
making the ``pitch-direction'' the binormal. With respect to the contour length the point
along the $0$-path gets shifted
by a shortening factor $\rho$, for which we will use its expectation value. The
deflection length in a confined channel is
considerably shorter than the persistence length of the chain. In general one
can expect, in conditions that allow for a perturbative expansion, that the
length
scales of the fluctuations are small compared to the global lengthscales.
The main assumption in the following is: the wavelength of thermal undulation is
considerably shorter than those of the writhing $0$ path. More precisely the
curvature and Fresnet torsion, which is $2\pi$ times the writhe density of the
plectoneme, are small compared to the wavenumbers of thermal undulations.
Neglecting contributions from the $0$-path torsion and curvature we arrive at
the following equations:
\begin{align}
\begin{split}
 \vec{r}(s)&:=\vec{r}_0(\rho s)+ u_i(s)\vec{t}_{\bot,0}^i(\rho
s)\qquad\Rightarrow \\
\vec{t}(s)&\simeq\rho\vec{t}_0(\rho
s)+\dot{u}_i(s)\vec{t}_{\bot,0}^i(\rho s)\qquad\Rightarrow \\
\vec{\dot{t}}(s)&\simeq\rho^2\vec{\dot{t}}_0(\rho
s)+\ddot{u}_i(s)\vec{t}_{\bot,0}^i(\rho s)
\end{split}
\end{align}
We conclude that we can treat the channel as being straight for thermal
fluctuations, provided we multiply the curvature of the $0$-chain by $\rho^2$.
The bending energy of the $0$ chain, being proportional to the curvature
squared, acquires then a factor of $\rho^4$.

For the writhe calculation we make again use of Fullers
equation~\ref{eq:fuller}, now with a homotopy from the $0$-path. It is
fairly easy to prove, using short intervals and continuity, that the projection
of the fluctuating path on the
$0$-path along the normal bundle forms a valid homotopy for Fullers
equation.
We write $\wri(s)=\wri_0(\rho s)+\Delta \wri(s)$. The $0$-path writhe is as
before but multiplied by $\rho$, since the tangent at $\rho s$ does not change,
but its rate of change does.

Finally applying Fullers equation results in:
\begin{align}
 \Delta\wri(s)&= \frac{1}{2\pi}\frac{(\vec{t}_0(\rho s)\wedge
\vec{t}(s))(\dot{\vec{t}}(s)+\rho\dot{\vec{t}}(\rho s))}{1+\rho }\simeq
\frac{1}{4\pi}(\dot{u}_r(s)\ddot{u}_p(s)-\ddot{u}_r(s)\dot{u}_p(s))
\end{align}
up to quadratic order and using the same assumptions as before.

\section{\label{ap:multi}Multi-plectoneme entropy}
In this appendix the number of configurations of a chain with a total
plectoneme length $\Lp$, divided over $m$ plectonemes, $Z_m(\Lc,\lk)$, is
calculated.
We make lengths dimensionless by rescaling them with a cutoff. The natural
cutoff is not a priori clear. One could argue for the
deflection length $\lambda$, which is the natural length-scale in the tails, or
alternatively for the \nm{3.5} helical repeat which must be a scale where
nucleation of loops are influenced by. We will choose the latter as the length scale for
positioning and length distribution of the plectonemes in our calculations.
Measurement data are, due to noise, not yet precise enough to differentiate
between possible length scales. Fluctuations in plectoneme length $L$ are mostly balanced by twist fluctuations. Their contribution to the partition sum is independent of the way $L$ is split between plectonemes, and thus:
\begin{align}
 Z_m(\Lc,\nu)=\int_0^{\Lc-m\Ll}\de L z_m(L)\exp(-\Lc\free_m(L,\nu))
\end{align}
with $z_m(L)$ the density of states at constant $L$. We assume that $z_m(L)\simeq z_m(\Lp)$, constant over the sharply peaked minimum of $\free_m(L,\nu)$ around $L=\Lp$. From Eqs.~\eqref{eq:Lpmult} and~\eqref{eq:ftwist} and because of the sharp minimum we find that the integral can be approximated by a Gaussian:
\begin{align}
 Z_m(\Lk)=\sqrt{\frac{\Lc}{2\pi\pc^{\rm{ren}}\rp^2\wri^2}}z_m(\Lp)\exp(-\Lc\free_m(\Lp,\Lk))
\end{align}

We first treat the case with hardcore interactions between the
plectonemes.
For a configuration with one plectoneme of length $\Lp$ and loop-length $\Ll$
the number of possible configurations is $\Lc-\Ll-\Lp$, the length along the
chain the plectoneme can end. In case of $2$ plectonemes sharing the
length $\Lp$, the first plectoneme we encounter, with plectoneme length
$\Lambda_1$, can have a position $x_1$ between $\Ll+\Lambda_1$ and $\Lc-\Ll-(\Lp-\Lambda_2)$, while the
second plectoneme can have a position $x_2$ in the interval
$[x_1+\Lambda_2+\Ll,\Lc]$. It is
easy to show using induction that the partition sum for $m$ loops can be written
as:
\begin{align}
 z_m^{\rm{hc}}(\Lp)=\prod_{i=1}^{m-1}\left(\int_0^{\Lp-\sum_{j=0}^{i-1}\Lambda_j
}\de \Lambda_i\right)
\prod_{k=1}^{m}\left(\int_{x_{k-1}+\Ll+\Lambda_{k}}^{\Lc-(m-k)\Ll-(\Lp-\sum_{
q=1}^{k}\Lambda_q)}\de x_k\right),
\end{align}
with $\Lambda_0=x_0=0$. To shorten the notation we define an effective chain
length $\Lc':=\Lc-m\Ll$.
The second product, which we denote by $y_m$, integrates over all positions of
the plectoneme. It can be written as
\begin{align}
 y_m(\Lc'-\Lp)&= \prod_{k=1}^{m}\left(\int_{0}^{\Lc'-\Lp-\sum_{q=0}^{k-1}x_q}\de
x_k\right)\nonumber\\
&= \int_0^{\Lc'-\Lp}\de x_1 y_{m-1}(\Lc'-\Lp-x_1) \nonumber \\
&= \mathcal{L}^{-1}\left(\frac{1}{t^{m+1}}\right)(\Lc'-\Lp) \\
&=\frac{(\Lc'-\Lp)^m}{m!}
\end{align}
where in the third step $\mathcal{L}^{-1}$ denotes an inverse Laplace transform
and the faltung theorem has been used.
The first term can be calculated analogously, resulting in the comprehensive
result:
\begin{align}\label{eq:hac1}
 Z_m^{\rm{hc}}(\Lc,\nu)&\sim\sqrt{\Lc}\frac{\Lp^{m-1}}{(m-1)!}\frac{(\Lc'-\Lp)^m}{m!}e^{-\Lc \free_m}
\end{align}

This hard core interaction is probably not entirely realistic. With a minor
penalty plectonemes can have some overlap. The effects of plectoneme
interactions come into play only when most of the free DNA has been used. As a
test the calculations can be performed with the other extreme of noninteracting
plectonemes. Defining $\Lc'':=\Lc-\Ll$, and again implicitly rescaling all
lengths by the helical repeat, we find as combinatorial factor:
\begin{align}\label{eq:ni}
 z_m^{\rm{ni}}(\Lp)&=\frac{1}{m!}\prod_{i=1}^m\left(\int_0^{\Lp-x_{i-1}}\de x_i
(\Lc''-x_i)\right)(\Lc''-\Lp+\sum_{i=1}^{m-1} x_i)\nonumber \\
&=\frac{1}{m!}\mathcal{L}^{-1}\left(\frac{\Lc''}{t}-\frac{1}{t^2}\right)^m(\Lp)\notag
\\
&= \frac{1}{m!}\sum_{k=0}^m\left(\begin{array}{c} m\\k\\
\end{array}\right)\frac{(-1)^k \Lc''^{m-k}\Lp^{m+k-1}}{(m+k-1)!},
\end{align}
which, not necessarily providing more clarity, can be written using a confluent
hypergeometric function as:
\begin{align}
 Z_m^{\rm{ni}}(\Lc,\nu)&\sim\sqrt\Lc\frac{(\Lc'')^m\Lp^{m-1}{}_1\!\textrm{F}_1(-m,m
,\Lp/\Lc'')}{m!(m-1)!}e^{-\Lc \free_m}
\end{align}

The experiments we have analyzed are mostly situated in the relatively low plectoneme density range, making the difference between
these two extremes too small to be able to tell how soft the plectoneme interaction is, especially in light of the fact that the relative writhe densities of plectoneme and loop do not differ much.

\section{\label{app:Lambert}Analyzing the plectoneme length and number density}
In this appendix we analyze the behavior of the plectoneme length and number density as the number of turns increases.
Note that Eq.~\eqref{eq:mextr} is symmetric under the transformation $r_n,r_{\wri}\rightarrow1-r_n,1-r_{\wri} $ whereby $\lp\leftrightarrow\bar{z}$.
The question is when to expect multiple plectonemes. Defining $\mu_l:=\mu/\lp$ and $\mu_z:=\mu/\bar{z}$, we can formally solve Eq.~\eqref{eq:mextr} exponentiating it as:
\begin{align}
 r_{\wri}\mu_l\exp(r_{\wri}\mu_l)=\exp(-(1-r_{\wri})\mu_z-\Delta')\frac{r_{\wri}}{\mu_z}
\end{align}
This is the defining equation for the Lambert W function~\cite{NIST:DLMF}. Since the right hand side is positive we need the principal branch $\lamp$ as only real valued branch:
\begin{subequations}
\begin{align}
 \mu_l&=\frac{1}{r_{\wri}}\lamp\left(\frac{r_{\wri}}{\mu_z}e^{-(1-r_{\wri})\mu_z-\Delta'}\right)\label{eq:lambertl}\\
\mu_z&=
\begin{cases}
\frac{1}{1-r_{\wri}}\lamp\left(\frac{(1-r_{\wri})}{\mu_l}e^{-r_{\wri}\mu_l-\Delta'}\right)& \text{if $r_{\wri}<1$ or $r_{\wri}>1\wedge \mu_z<\frac{1}{r_{\wri}-1}$}\\
\frac{1}{1-r_{\wri}}\lamm\left(\frac{(1-r_{\wri})}{\mu_l}e^{-r_{\wri}\mu_l-\Delta'}\right)& \text{if $r_{\wri}>1\wedge \mu_z>\frac{1}{r_{\wri}-1}$}\\
\frac{1}{\mu_l}e^{-\mu_l-\Delta'} & \text{if $r_{\wri}=1$}.
\end{cases}\label{eq:lambertz}
\end{align}
\end{subequations}
The second equation follows from the first by symmetry, but care has to be taken which branch to follow. The argument is in this case negative and a second real branch exists, $\lamm$. The crossover happens when the argument reaches its minimum of $-1/e$.

 We next  analyze the scaling behavior at the end of the plectoneme slope, when $\bar{z}$ reaches zero, using Eq.~\eqref{eq:lambertz}.
\begin{enumerate}
 \item Case $r_{\wri}<1$: Suppose we end up with a finite density of plectonemes then $\mu_z\rightarrow \infty$ and so the argument of $\lamp\rightarrow\infty$ in Eq.~\eqref{eq:lambertz}. As this is impossible, we have $\mu\rightarrow 0$ and $\lp\rightarrow 1$. Since $\mu_l$ tends to zero the argument of $\lamp$ goes to infinity. To lowest order we find:
\begin{align}
\begin{split}
 \lim_{\bar{z} \to 0}\mu_z&\simeq\frac{1}{1-r_{\wri}}\log\left[\frac{1-r_{\wri}}{\mu_l}e^{-(1-r_{\wri})\mu_l-\Delta'}\right]\simeq-\frac{1}{1-r_{\wri}}\log(\mu)\Rightarrow\\
\lim_{\bar{z} \to 0}\mu&\simeq\frac{\bar{z}}{1-rw}\lamp(\frac{1-rw}{\bar{z}})\simeq -\frac{\bar{z}\log(\bar{z})}{1-rw}
\end{split}
\end{align}
 \item Case $r_{\wri}>1$: The density of plectonemes can not be zero when $\bar{z}\downarrow 0$, since then $\lp$ has to be one and the argument would dive below $-1/e$ where the Lambert function is not real. So $\mu_z\rightarrow\infty$ and thus goes the argument of $\lamm$ to zero, from which follows that $\lp\downarrow 0$ and consequently $\mu\uparrow 1$. Here we find as asymptotic
$$\lim_{\bar{z} \to 0}\lp\simeq \frac{r_{\wri}}{r_{\wri}-1}\bar{z} $$
 \item Case $r_{\wri}=1$: Now the plectoneme number density goes to zero as $\mu\sim \sqrt{\bar{z}}e^{-\Delta'/2}$
\end{enumerate}
Note that these two opposite limits at vanishing extension do not depend on $\Delta$ or its renormalized primed version.

These expressions are useful to obtain a parametrization of the $\mu, \lp$ or $\bar{z}$ curves as function of $r_n$ or $\nu$. Since
\begin{align}\label{eq:parametrization}
 \mu&=\frac{\mu_z\mu_l}{\mu_z\mu_l+\mu_z+\mu_l} & r_n&= \frac{(1+r_{\wri}\mu_l)\mu_z}{\mu_z\mu_l+\mu_z+\mu_l}
\end{align}
we can use this and Eq.~\eqref{eq:lambertl} as a parametrization of the $\mu(r_n)$ curve by $\mu_z$. At the start of the slope $\mu_z$ is zero. We just need to know its value at the end of the slope, which  follows from the previous scaling relations as being infinite.
From Eq.~\eqref{eq:parametrization} we then obtain that the maximum value $r_n$ reaches:
\begin{align}
 \lim_{\mu_z\to \infty}r_n&=\begin{cases}
				1 & \text{if $r_{\wri}\leq1$}\\
				r_{\wri}& \text{if $r_{\wri}>1$}
                           \end{cases}
\end{align}
which just tells us how much linking number the chain can absorb. More important is that if one wishes to interpret the turns extension slopes in terms of a single plectoneme then when $r_{\wri}>1$ the plectoneme parameters have to be adjusted. Some plots generated by this parametrization are in Fig.~\ref{fig:multiplot}.

Of practical importance is to know when multiple plectonemes become significant not far after the transition, since it is there where most measurements are performed.
Single plectoneme behavior we expect when most additional linking number goes into a growing plectoneme, or:
\begin{subequations}
\begin{align}\label{eq:mpfact1}
 \lp\simeq r_n \Rightarrow \lp \gg r_{\wri}\mu
\end{align}
From Eq.~\eqref{eq:lambertl} and the properties of $\lamp$ it follows that  close to the transition, where $\bar{z}\simeq1$:
\begin{align}\label{eq:mpfact2}
 \frac{r_{\wri}}{\mu_z}e^{-\Delta'}\ll1 \Rightarrow \mu \gg r_{\wri}e^{-\Delta'}
\end{align}
Finally combining Eqs.~\eqref{eq:mpfact1} and~\eqref{eq:mpfact2} results in the following inequality a mostly single plectoneme configuration should abide to:
\begin{align}
 \lp\gg \zeta:=r_{\wri}^2e^{-\Delta'}
\end{align}
 \end{subequations}
Since close to the transition side of the turn extension slope $\lp\ll1$, the \emph{multi plectoneme} factor, $\zeta$, is an indicator for the appearance of several plectonemes. Once $\zeta$ becomes of the order one, the change in the number of plectonemes plays an important part in the conversion of added linking number into writhe. Fig.~\ref{fig:mpfac} shows the resulting  $\zeta$ over a range of salt concentrations and forces.

The maximum of the loop density over the full range of allowed linking number densities is also straightforward to calculate to lowest order:
\begin{align}\label{eq:mumax}
\begin{split}
 \mu_{\rm{max}}\simeq& \\
e^{-\Delta'/4}&\frac{\cosh(\frac{\Delta'}{4})-r_{\wri}e^{-\Delta'/4}}{\sinh(\frac{\Delta'}{4})+(1-r_{\wri})r_{\wri}e^{-\Delta'/4}}\left[r_{\wri}e^{-\Delta'/4}+\left(1-\sqrt{\frac{\sinh(\frac{\Delta'}{4})+r_{\wri}e^{-\Delta'/4}}{\cosh(\frac{\Delta'}{4})-r_{\wri}e^{-\Delta'/4}}}\right)\sinh(\frac{\Delta'}{4})\right]
\end{split}
\end{align}
\bibliographystyle{apsrev4-1}
\bibliography{longplect}
\end{document}